\documentclass[prc,aps,floatfix,showpacs,twocolumn,nofootinbib]{revtex4-2}
\usepackage{dcolumn}
\usepackage{slashed}
\usepackage{bm}
\usepackage{xcolor}
\usepackage[markup=underlined]{changes}
\usepackage{graphicx}
\usepackage{amssymb,amsmath}
\usepackage[utf8]{inputenc} 							
\usepackage{bbm}
\usepackage{simplewick}
\usepackage{float}
\usepackage{simplewick}

\def\tri{{{}^3{\rm H}}}
\def\het{{{}^3{\rm He}}}
\def\heq{{{}^4{\rm He}}}

\def\beo{{{}^8{\rm Be}}}
\def\cdo{{{}^{12}{\rm C}}}
\def\bun{{{}^{11}{\rm B}}}

\def\bmr{{\bm r}}

\def\bmx{{\bm x}}

\def\bmp{{\bm p}}
\def\bmk{{\bm k}}
\def\bmq{{\bm q}}
\def\bmj{{\bm j}}
\def\bme{{\bm e}}

\def\bmR{\bm R}

\def\bmP{{\bm P}}

\def\y{{\bm y}}
\newcommand{\bmsi}{{\bm \sigma}}

\definechangesauthor[color=red]{Michele}
\definechangesauthor[color=purple]{Elena}
\definechangesauthor[color=blue]{Luca}
\definechangesauthor[color=brown]{Ale}
\definechangesauthor[color=green]{Laura}
\definechangesauthor[color=orange]{Carlo}

\begin{document}
\title{The X17 boson and the $d(p,e^+ e^-)\het$ and $d(n,e^+ e^-)\tri$ processes: a theoretical analysis}

\author{M. Viviani$^1$, E. Filandri$^{2,1}$, L. Girlanda$^{3,4}$, C. Gustavino$^5$, A. Kievsky$^1$, and
L.E. Marcucci$^{2,1}$}

\affiliation{
$^1$INFN-Pisa, I-56127, Pisa, Italy \\
$^2$Department of Physics ``E. Fermi'', University of Pisa, I-56127 Pisa, Italy \\  
$^3$Department of Mathematics and Physics, University of Salento, I-73100 Lecce, Italy \\
$^4$INFN-Lecce, I-73100 Lecce, Italy \\  
$^5$INFN Sezione di Roma, 00185 Rome, Italy}

\begin{abstract}
The present work deals with the $e^+$-$e^-$ pair production in the
$d(p,e^+ e^-)\het$ and $d(n,e^+ e^-)\tri$ processes, 
in order to make evident possible effects due to the exchange of a hypothetical low-mass boson,
the so-called X17.  These processes are studied for energies of the incident
beams in the range 18-30 MeV, in order to have a sufficient energy to produce such a boson,
whose mass is estimated to be around 17 MeV. 
We first analyze them  as a purely electromagnetic processes,
in the context of a state-of-the-art approach
to nuclear strong-interaction dynamics and nuclear electromagnetic currents,
derived from chiral effective field theory ($\chi$EFT).  Next, we examine
how the exchange of a hypothetical low-mass boson would impact the cross sections
for such processes. 
We consider several possibilities, that this boson is either a scalar, pseudoscalar, vector, or
axial particle. The main aim of the study is to exploit the specular structure of the $^3$He and $^3$H nuclei to investigate the isospin dependency of the X17-nucleon interaction, as the alleged ``proto-phobicity''.
\end{abstract}


\maketitle

\section{Introduction}
\label{sec:intro}
In the last years, there were claims~\cite{Krasznahorkay:2015iga,Krasznahorkay:2019lyl,Krasznahorkay:2021joi,Sas:2022pgm,Krasznahorkay:2022pxs}
that an unknown
particle (denoted as ``X17'') had been observed in the processes $^7$Li$(p,e^+ e^-)^8$Be,
$\tri(p, e^+ e^- )\heq$, and $\bun(p, e^+ e^- )\cdo$ at the ATOMKI experimental facility situated in Debrecen (Hungary).
The $\beo$ result has been confirmed very recently by another experiment performed at the VNU University of Science in Hanoi (Vietnam)~\cite{Anh:2024req}.
These claims were based on a $\approx 7 \sigma$ excess of events in the
angular distribution of leptonic pairs produced in these reactions, which have
a $Q$-value of about $20$ MeV.  The excess could be explained by positing the emission of an unknown
boson with a mass of about $17$ MeV decaying into $e^+ e^-$ pairs.

The possible existence of a new kind of low mass particle (at the MeV scale)
is a problem of current and intense theoretical and experimental interest
(see, for example, Ref.~\cite{Battaglieri:2017aum} and references therein).
 This interest is, in fact, part of a broader effort aimed at identifying dark matter
(DM).  The search for bosonic DM candidates had already started by several
years, by attempting to establish the possible existence of additional forces
(beyond gravity), mediated by these bosons~\cite{Pospelov:2008aaa} between
DM and visible matter.  To one such class of particles belongs the so-called
``dark-photon'', namely a boson of mass $M_{X}$ having the same
quantum numbers as the photon, and interacting with a Standard Model (SM)
fermion $f$ with a coupling constant given by $\varepsilon \,q_f$, where $q_f$ is the fermion
electric charge. Following several years of experimental searches,
``exclusion plots'' in the $\varepsilon$-$M_{X}$  parameter space were produced,
restricting more and more the allowed region~\cite{Battaglieri:2017aum,Banerjee:2018vgk}.
One of the most stringent limits was provided by the NA48/2 experiment~\cite{Batley:2015lha}.

The observation of the $\beo$, $\heq$, and $\cdo$ anomalies by the ATOMKI and VNU groups
soon spurred several theoretical studies.  In Ref.~\cite{Feng:2016jff}, 
the possibility that the X17 could be a vector boson
was investigated in detail.  In order to circumvent the NA48/2 limit, 
it was conjectured that the X17 could be ``proto-phobic'', namely that it would couple much more
weakly to the proton than to the neutron~\cite{Feng:2016jff}. 
Many other theoretical studies were published afterwards~\cite{Zhang:2017zap,Ellwanger:2016wfe,Fornal:2017msy,Dror:2017ehi,DelleRose:2017xil,DelleRose:2018eic,DelleRose:2018pgm,Alves:2017avw,Alves:2020xhf,Bordes:2019wcp,Nam:2019osu,Kirpichnikov:2020tcf,Fayet:2020bmb,Hayes:2021hin,Wong:2022kyg,Barducci:2022lqd,Hostert:2023tkg}. 

On the experimental side, there are several experiments (MEGII~\cite{Baldini:2018nnn}, DarkLight~\cite{Balewski:2014pxa},
SHiP~\cite{SHiP:2020noy}, the ``Montreal X-17 Project''~\cite{Azuelos:2022abcd},  TREK/E36~\cite{Dongwi:2022abcd},
PADME~\cite{Darme:2022zfw}, New JEDI~\cite{Bastin:2023utm}, MESA~\cite{Doria:2019sux},
and others~\cite{Battaglieri:2017aum}) planning specifically to search for such a light boson.  In addition, large
collaborations, such as BelleII~\cite{Kou:2018nap}, NA64~\cite{Banerjee:2020bbb}, and others, 
are dedicating part of their efforts in an attempt to clarify this issue.
Recently, the reaction $\het(n, e^+ e^- )\heq$ has been proposed as a possible new
process where the presence of X17 can be observed and, eventually, its properties can be
determined~\cite{Cisbani:2021x17}.

Other stringent experimental constraints on the possible magnitude of the X17 coupling constants
with nucleons come from various pion-decay experiments (as the KTeV anomaly~\cite{KTeV:2006pwx}
and SINDRUM-I experiment~\cite{SINDRUM:1989qan}). See the analyses of Refs.~\cite{Barducci:2022lqd,Hostert:2023tkg} for 
a comprehensive discussion.
Most of the analyses performed so far are based on the hypothesis that the X17 production takes place as a two-step
process: {\it i)} the formation of a specific resonance with well defined quantum numbers of the final nucleus; {\it ii)}
the emission of the X17 in the transition from this resonant state to the ground state. However, the studies performed
assuming a full dynamic initial scattering state~\cite{Viviani:2021stx,Gysbers:2023wug} have shown that the situation is 
not so simple, as the contribution of many resonances and even of the direct capture from $P$ waves
have to be taken into account. Therefore, we stress the necessity of studying processes where the nuclear dynamics
is fully taken into account. 

Here, we present a study of the  $d(p, e^+ e^- )\het$
and  $d(n, e^+ e^- )\tri$ reactions. The advantage of studying these processes is that for them it is possible to perform
accurate {\it ab initio} calculations to describe bound- and continuum-states.
We use the hyperspherical-harmonics (HH) method to determine them~\cite{Kievsky:2008es,Marcucci:2020fip},
by taking fully into account the three-body dynamics.
The employed nuclear Hamiltonians are very accurate. First of all we have employed the phenomenological Argonne
$v_{18}$ (AV18)~\cite{Wiringa:1994wb} two-nucleon ($2N$) and the Urbana IX (UIX)~\cite{Pudliner:1995wk}
three-nucleon ($3N$) interaction. This interaction will be denoted hereafter as AV18UIX.
Furthermore, we have also considered the Norfolk $V_{Ia}$ 2N interaction~\cite{Piarulli:2016vel,Piarulli:2017dwd}
plus the 3N force~\cite{Epelbaum:2002vt} derived within the framework of chiral effective field theory ($\chi$EFT).
This interaction will be denoted in the following as NVIa3N.
To perform this study, we need accurate accompanying electromagnetic (EM) currents, as well.
We have used those derived from the $\chi$EFT  study of
Refs.~\cite{Pastore:2008ui,Pastore:2009is,Pastore:2011ip,Piarulli:2012bn,Schiavilla:2018udt}.
The LECs entering these EM currents have been fixed by reproducing the trinucleon magnetic moments
for both AV18UIX and NVIa3N interactions. 

Regarding the X17 interaction, following Ref.~\cite{Viviani:2021stx}, we consider the general case of a Yukawa-like interaction
between this boson and a SM fermion of species $f$ (specifically, $u$ and $d$ quarks, and electrons) with the
coupling constant expressed as $\varepsilon_f \,e$, where $e>0$ is
the unit electric charge.  The X17 boson must decay promptly in $e^+$-$e^-$ pairs for these to be detected
inside the experimental setup.  This observation actually introduces a {\it lower limit} to the possible
values of $\varepsilon_e$, the X17-electron coupling constant.  These limits are also established by various electron
beam-dump experiments (see, for example, Ref.~\cite{PhysRevD.86.095019} and
references therein).  For $M_X\approx17$ MeV, the most stringent lower bound, $|\varepsilon_e| > 2 \times 10^{-4}$, comes
from the SLAC E141 experiment~\cite{PhysRevLett.59.755}, while the upper bound $|\varepsilon_e|< 2 \times 10^{-3}$ has been
set by the KLOE-2 experiment~\cite{Anastasi:2015qla}. However, we will show that in the considered
nuclear processes there is no sensitivity to $\varepsilon_e$. In the following, we will assume
$\varepsilon_e=10^{-3}$ and study how the angular distribution of the two leptons
is affected by the values of the spin and parity of the X17 boson.

The Lagrangian interaction terms with hadrons (specifically, nucleons and pions) are obtained
by placing the X17 as an external source in the QCD Lagrangian~\cite{Weinberg:1968de,Gasser:1983yg,Bernard:1995dp}.
Considering only interactions invariant under parity, charge conjugation and time reversal,
we examine four possible X17-quarks interaction types: scalar, pseudoscalar, vector, and axial.
These Lagrangian terms will depend on a number of low-energy constants (LECs), which take into account
the hadron dynamics induced by QCD. Many of these LECs are known as they also enter various
nuclear processes, as for example nucleon-nucleon scattering, pion decay, etc. Therefore, they can be
extracted from experimental data. Then, using the quark-X17 coupling constants given in
Ref.~\cite{Viviani:2021stx}, as determined in order to reproduce the $p(\tri,e^+e^-)\heq$ experimental data
of Ref.~\cite{Krasznahorkay:2019lyl}, we will be able to predict the contribution of an X17 (of mass $17$ MeV)
in the  $d(p, e^+ e^- )\het$ and $d(n, e^+ e^- )\tri$) reactions. This can be
done for the four considered X17-quarks interaction types.

Assuming standard physics, the excited $^3$H and $^3$He nuclei have no resonant levels and decay to the ground state mainly
with the emission of a single photon via an electric dipole transition. The production of the X17 is possible only
above a well definite energy of the beam, $E_p>17.3$ MeV ($E_n>16.1$ MeV). Under such thresholds, no
peak should be observed. Also the position of peak in the $e^+-e^-$ angular distribution is strictly connected to
the beam energy and X17 mass value. Therefore, having the possibility to vary the beam energy, let us say in the $18-30$ MeV range,
one could put a severe constraint on the existence and mass of X17 (and whether it is either a scalar, vector, pseudoscalar, or axial particle). 
Moreover, as stated before, the possibility to compare the
$d(p, e^+ e^- )\het$ and $d(n, e^+ e^- )\tri$ experimental cross sections could give important information
regarding the isospin dependence of the X17 interaction with quarks. The experimental study of these
reactions has been already considered, see for example~\cite{universe10070285,Gustavino:2024wgb}.

The work described in this contribution is based on the study presented in Ref.~\cite{Viviani:2021stx}. 
In Sec.~\ref{sec:theo}, a brief description of the theoretical formalism is given. Then, in Sec.~\ref{sec:res}
the results of the calculations are reported and discussed.
Finally, in the last section, the perspectives of this study are given.

\section{Theoretical analysis}
\label{sec:theo}

\subsection{The interaction Lagrangian}
\label{sec:lag}
In this work, the interaction Lagrangian density at energy scale $\Lambda_H\sim 1$ GeV
is considered to be
\begin{equation}
  {\cal L}^c_{X}(x)=e\, \varepsilon^c_e\,\overline{e}(x)\, \Gamma^{c}\, e(x)\, X_{c}(x)+ {\cal L}^c_{q,X}(x)\ , \label{eq:LeX}
\end{equation}
where  $e(x)$ is the electron field and $X_{c}(x)$ the X17 field (see below). In the following, we will consider
four cases: $c=S,P,V,A$ for a scalar, pseudoscalar, vector, or axial boson. Correspondingly,
\begin{equation}
\Gamma^{c=S,P,V,A}=1, i\, \gamma^5 , \gamma^\mu, \gamma^\mu \, \gamma^5 \ .
\end{equation}
The various coupling constants will be always written in units of the electric charge $e>0$ ($e^2\,$=$\,4\pi\alpha$,
where $\alpha\approx 1/137$ is the fine structure constant).

The part ${\cal L}^c_{q,X}(x)$ describes the interaction of the X17 with quarks. For a scalar ($S$) or
pseudoscalar ($P$) X17 boson, we take as
\begin{equation}
  {\cal L}^c_{q,X}(x)=e\,{1\over\Lambda} \sum_{f=u,d,\ldots} m_f \varepsilon^c_f\,\overline{f}(x)\, \Gamma^{c}\, f(x)\, X_{c}(x)
  \ ,\quad c=S,P\ , \label{eq:LqX1}
\end{equation}
where $f(x)$ is the field of the quark of flavour $f$, $\Lambda$  an unknown high-energy mass scale,
and we have introduced explicitly the quark masses $m_f$, $f=u,d,\ldots$, in order to have  renormalization-scale invariant
amplitudes. In Eq.~(\ref{eq:LqX1}), the sum runs over the lightest fermions of the SM. Reducing to the case of
only $u$ and $d$ quarks, it is possible to rewrite this Lagrangian in terms
of the isodoublet quark fields $q(x)$, defined as
\begin{equation}
q(x)=\left[\!\begin{array}{c} 
         u(x)\\
         d(x)
         \end{array}\!\right]\ ,
\end{equation}
in the following way
\begin{equation}
 {\cal L}^{c}_{q,X}(x)= e\,{m_q\over\Lambda_S}\, \overline{q}(x)
    (\varepsilon^c_0 + \varepsilon^c_z \tau_3)\Gamma^c \, q(x)\, X_c(x)\ ,\quad c=S,P\ ,   \label{eq:Lqs2}
\end{equation}
where $\tau_3$ is a Pauli matrix, $m_q$ is the average light-quark mass, and we have introduced the coupling constants (again $c=S,P$)
\begin{eqnarray}
  \varepsilon^c_0 &=& {\Lambda_S\over\Lambda}\, {m_u \,\varepsilon^c_u +
   m_d \,\epsilon_d^c\over 2\,m_q}\ , \label{eq:eps_p_s}\\
  \varepsilon^c_z &=& {\Lambda_S\over\Lambda}\, {m_u \, \varepsilon^c_u - m_d\, \varepsilon^c_d\over 2\,m_q}\ , \label{eq:eps_m_s}
\end{eqnarray}
and a new scale $\Lambda_S$ which we set (arbitrarily) at 1 GeV. 

For a vector ($V$) or axial ($A$) X17, the Lagrangian is taken as
\begin{equation}
  {\cal L}^c_{q,X}(x)=e\, \sum_{f=u,d,\ldots} \varepsilon^c_f\,\overline{f}(x)\, \Gamma^{c}\, f(x)\, X_{c}(x)
  \ ,\quad c=V,A. \label{eq:LqX2}
\end{equation}
Therefore,  in case of only two quarks, it can be rewritten as
\begin{equation}
 {\cal L}^{c}_{q,X}(x)= e\,\, \overline{q}(x)
    (\varepsilon^c_0 + \varepsilon^c_z \tau_3)\Gamma^c\, q(x)\, X_c(x)\ ,  \label{eq:Lqv2}
\end{equation}
where we have introduced the coupling constants
\begin{eqnarray}
  \varepsilon_0^c &=&  {\varepsilon_u^c +
   \epsilon_d^c\over 2}\ , \label{eq:eps_s_v}\\
  \varepsilon_z^c &=&  {\varepsilon_u^c - \varepsilon_d^c\over 2}\ . \label{eq:eps_v_v}
\end{eqnarray}
Note that a proto-phobic X17 is defined to have $2\varepsilon_u^c+\varepsilon_d^c=0$, or equivalently
\begin{equation}
  3\varepsilon_0^c+\varepsilon_z^c=0\ .\label{eq:protophobic}
\end{equation}
Finally,  for the $c\,$=$\,S, P$ cases, the X17 field is a scalar field, $X_{c}(x)\,$=$\,X(x)$. On the other hand,
for the $c\,$=$\,V, A$ cases, the X17 field is a vector field, $X_{c}(x)\,$=$\,X_\mu(x)$.

Starting from these Lagrangians, it is possible to derive nucleon-X17 interaction Lagrangian densities in the framework
of $\chi$EFT. By retaining only leading-order contributions (and selected subleading ones in
the vector and pseudoscalar cases), one obtains (for a detailed derivation, see Ref.~\cite{Viviani:2021stx})
\begin{eqnarray}
\label{eq:e4s}
{\cal L}^{S}_{X}(x)\!&=&\!e\, \overline{N}(x)[\eta^{S}_0 +\eta^{S}_z\, \tau_{3}]N(x)\,X(x) \ ,\\
\label{eq:e4p}
{\cal L}^{P}_{X}(x)\!&=&\!e\, \eta_z^{P}\, \pi_3 (x)\, X(x) + e \, \eta^{P}_0 \overline{N}(x)i\,\gamma^5N(x)\,X(x)\ ,\\
\label{eq:e4v}
{\cal L}^{V}_{X}(x)\!&=&\!e\,  \overline{N}(x)[\eta^{V}_0 +\eta^{V}_z\, \tau_{3}]\gamma^\mu\,N(x)\,X_\mu(x)  \\
&&\!+\frac{e}{4\, m_N}\,  \overline{N}(x)[\kappa_0 \eta^{V}_0 +\kappa_z\eta^{V}_z\tau_{3}]\sigma^{\mu\nu}\,N(x) F^X_{\mu\nu}(x)  \ ,\nonumber\\
{\cal L}^{A}_{X}(x)&=&\!e\,  \overline{N}(x)[\eta^{A}_0 +\eta^{A}_z\, \tau_{3}]\gamma^\mu\gamma^5\,N(x)\,X_\mu(x)\ ,
\label{eq:e4a}
\end{eqnarray}
where $m_N$ is the nucleon mass, $N(x)$ is the iso-doublet of nucleon fields, $\pi_3(x)$ is the third component of
the triplet of pion fields, and $F^X_{\mu\nu}(x)\,$=$\,\partial_\mu\, X_\nu(x)-\partial_\nu\, X_\mu(x)$
is the X17 field tensor. The hadron-X17 coupling constants $\eta^c_0$ and $\eta^c_z$ are linear
combinations of the quark-X17 coupling constants $\varepsilon_u^c$ and $\varepsilon_d^c$~\cite{Viviani:2021stx}
\begin{eqnarray}
\eta_0^{S} &=& - {4 m_\pi^2 c_1\over  \Lambda_S} \varepsilon^S_0 \label{eq:csp}\ ,\nonumber \\
\eta_z^{S} &=& -{2 m_\pi^2 c_5\over \Lambda_S} \varepsilon^S_z \label{eq:csm}\ , \nonumber \\
\eta_0^{P} &=& 2 {m_\pi^2 \,m_N (d_{18}+2\,d_{19})\over \Lambda_S}\varepsilon^P_0 \ , \label{eq:cpp}\nonumber\\
\eta_z^{P} &=& {m_\pi^2 f_\pi \over  \Lambda_S} \varepsilon^P_z \ , \label{eq:cpm}\nonumber\\
\eta_0^{V} &=& 3 \varepsilon^V_0\ , \label{eq:cvp}\\
\eta_z^{V} &=& \varepsilon_z^V ,\qquad  \label{eq:cvm}\nonumber\\
\eta_0^{A} &=& (3\,F-D)\varepsilon^A_0\ , \label{eq:cap}\nonumber\\
\eta_z^{A} &=& (F+D)\varepsilon_z^A\ , \nonumber\label{eq:cam}
\end{eqnarray}
where $c_1$, $c_5$, $F$, etc., are LECs
entering the nuclear chiral Lagrangian.
In the vector case, we have included also the subleading term proportional to
$F^X_{\mu\nu}$ and where 
\begin{equation}
    \kappa_0\,=\,\kappa_p+\kappa_n\ , \qquad
    \kappa_z\,=\,\kappa_p-\kappa_n\ , \label{eq:kappa}
\end{equation}
$\kappa_p$ and $\kappa_n$ being the anomalous
magnetic moments of the proton and neutron, respectively. 
In the pseudoscalar case, the interaction at leading order in the power counting 
originates from the direct coupling of the X17 to the pion.  However, since
the associated coupling constant is expected to be suppressed~\cite{Alves:2017avw,Alves:2020xhf},
we have also considered an isoscalar coupling of the X17 to the nucleon, even though
it is subleading, at least nominally, in the $\chi$EFT power counting relative
to the isovector one.  As per the axial case,
the leading order ${\cal L}^{A}_{X}(x)$ contains an additional term  
of the form $\partial^\mu \pi_3 (x)\, X_\mu(x)$, which we have dropped. In fact, this
term leads to a X17-nucleon current proportional to $q^\mu/m_\pi^2$ (for low momentum
transfers) which, when contracted with the lepton axial current, produces a contribution proportional to $(m_e/m_\pi)^2$,
and hence negligible when compared to that resulting from the X17 direct coupling to the nucleon.

The X17-induced nuclear current by each of the (leading order)
Lagrangians in Eqs.~(\ref{eq:e4s})--(\ref{eq:e4a}) can be easily calculated, for example,
in time-ordered perturbation theory. For simplicitly,
in this work we include the one-body contributions only,
which also coincides with the leading order in the chiral expansion
of the various amplitudes, see also Ref.~\cite{Viviani:2021stx}. As an example, the
scalar case is detailed in Appendix~\ref{app:a}. We have only retained the
leading-order terms in the non-relativistic expansion of the various
amplitudes. The nuclear currents $J^c_X$ are simply given as:
\begin{eqnarray}
J_X^S&=& \eta^S_0\, \rho^{S+} +\eta_z^S\, \rho^{S-}\ ,\label{eq:JXS}\\
J_X^P&=& {g_A\over 2 f_\pi}\eta_z^P\,\frac{q}{q^2+m_\pi^2}\, {\rho}^{P-}  + \eta_0^P\, \frac{q}{2 m_N} \, {\rho}^{P+}\ ,\label{eq:JXP}\\
\rho_X^V&=& \eta^V_0\, \rho^{S+} +\eta_z^V\, \rho^{S-}\ ,\label{eq:JXV1}\\
{\bm J}^{V}_X&=& \eta_0^V\left( {\bm j}^{V+}+\kappa_0\, \overline{\bm j}^{V+}\right)
+\eta_z^V\left( {\bm j}^{V-}+\kappa_z\, \overline{\bm j}^{V-}\right) \ , \label{eq:JXV2}\\
\rho_X^A&=& \eta^A_0\, \rho^{A+} +\eta_z^V\, \rho^{A-}\ ,\label{eq:JXA1}\\
{\bm J}^{A}_X&=& \eta_0^A {\bm j}^{A+}+\eta_z^A {\bm j}^{A-}\ ,\label{eq:JXA2}
\end{eqnarray}
where in the $c=V$, $A$ case the current has both a time (denoted as $\rho^c_X$) and a space (denoted as ${\bm J}^c_X$) component.
The nuclear currents above have been written in terms of the following single-particle ``basic'' adimensional
operators:
\begin{eqnarray}
\rho^{S\lambda}(\bmq)&=&\sum_{i=1}^A e^{i\bmq\cdot\bmr_i}\, P^{\lambda}_i \ ,\label{eq:e94}\\ 
\rho^{P\lambda}(\bmq)&=&\sum_{i=1}^A e^{i\bmq\cdot\bmr_i}\,i\,\hat{\bm q}\cdot \bmsi_i \,P^{\lambda}_i \ ,\label{eq:e95} \\
 \rho^{A\lambda}(\bmq)&=&\sum_{i=1}^A  \frac{1}{2\, m_N} \, \left[ e^{i\bmq
 \cdot\bmr_i}\, , \,\bmp_i\cdot\bmsi_i \right ]_+ \,P^{\lambda}_i  \ ,\label{eq:e96} \\
 {\bm j}^{V\lambda}(\bmq)&=&\sum_{i=1}^A\frac{1}{2\,m_N} \left[ e^{i\bmq\cdot\bmr_i}\, , \,\bmp_i \right ]_+ P^{\lambda}_i\ ,\label{eq:e97}\\
  \overline{{\bm j}}^{V\lambda}(\bmq) &=&  \sum_{i=1}^A\frac{i}{2\,m_N} \,e^{i\bmq\cdot\bmr_i}\, \bmq\times \bmsi_i \, P^{\lambda}_i  \ ,\label{eq:e98}\\
{\bm j}^{A\lambda}(\bmq)&=&\sum_{i=1}^A e^{i\bmq\cdot\bmr_i}\, {\bm \sigma}_i\,P^{\lambda}_i \ ,
\label{eq:e99}
\end{eqnarray}
where $\lambda=\pm$ with $P^+_i\,$=$\,1$ and $P^-_i\,$=$\,\tau_{i,3}$,
${\bm p}_i$ is the momentum operator,
and $[\cdots  ]_+$ denotes the anticommutator.
In the expressions above, the distances $\bmr_i$ are
 relative to the center-of-mass position ${\bmR_{CM}}$ of the $A=3$ nucleons.
 In the matrix elements, the integration over ${\bmR_{CM}}$
 gives the momentum conservation $\delta$-function.

Starting from the expressions above, at first order in perturbation theory, the amplitude
for the emission of an $e^+$-$e^-$ pair between nuclear initial and final states is written
in general as 
\begin{equation}
   T_{fi}=T_{fi}^{EM}+T_{fi}^{cX}\ , \label{eq:ampli1}
\end{equation}
where $T_{fi}^{EM}$ is the one-photon-exchange amplitude written as
\begin{eqnarray}
  T_{fi}^{EM}&=&4\pi\alpha\, \frac{(\overline{u}_-\,\gamma_\mu \,v_+)\, j^{\mu}_{EM}}
  {q^\mu\, q_\mu}\ ,   \label{eq:e1} \\
  j^{\mu}_{EM}&=& \langle\Phi_{m_3}| J^{\mu\dagger}_{EM}| \Psi_{m_2,m_1}\rangle\ , \label{eq:e1b}
\end{eqnarray}
where $J^\mu_{EM}$ is the nuclear EM current operator.
Above, $\alpha$ is the fine structure constant, $q^\mu$ is the four-momentum transfer
defined as the sum of the outgoing-lepton four momenta, and 
$u_-$ and $v_+$ are, respectively, the electron and positron spinors.
Moreover, $\Psi_{m_2,m_1}$ is the initial wave function describing the $N+d$ scattering state (see below),
while $\Phi_{m_3}$ is the trinucleon ground state wave function of spin projection $m_3$. 

The term $T_{fi}^{cX}$ is the amplitude with the exchange of a unknown boson X17 of type $c=S,P,V,A$, namely
\begin{eqnarray}
  T_{fi}^{cX}&=&4\pi\alpha \, \frac{\varepsilon_e\,(\overline{u}_-\,\Gamma_{c} \,
    v_+)\, j_{X}^{c}}{q^\mu\, q_\mu-M_X^2}\ ,   \label{eq:e2}\\
  j^{c}_{X}&=& \langle\Phi_{m_3}| J^{c\dagger}_{X}| \Psi_{m_2,m_1}\rangle\ , \label{eq:e2b}  
\end{eqnarray} 
where $M_X$ is the mass of the X17 particle, and $j^{c}_{X}$ represents the matrix element of
the X17-induced nuclear current $J^{c\dagger}_{X}$ given in Eqs.~(\ref{eq:JXS})--(\ref{eq:JXA2}),
which depends on the X17-hadron coupling constants $\eta^c_0$ and $\eta^c_z$. Note that in the case
$c=V$, $A$, such an operator is a four-vector. 

\subsection{$N-d$ wave functions and kinematics of the reactions}
\label{sec:kine}
In the laboratory frame, the initial state consists of an incoming proton 
or neutron of momentum ${\bmp}$ and spin projection $m_1$, and a
bound deuteron in spin state $m_2$ at rest.  Its wave function $ \Psi_{m_2,m_1}$ is
such that, in the asymptotic region of large separation ${\bm y}_\ell$ between the isolated nucleon (particle $\ell$) and
the deuteron (particles $ij$), it reduces to
\begin{equation}
\label{eq:e11}
\Psi_{m_2,m_1}(\bmp) \longrightarrow 
{1\over\sqrt{3}} \sum_{\ell=1}^3 \phi_{m_2}(ij) \chi_{m_1}(\ell) \, \Phi_{\bmp}(\y_\ell) \ ,
\end{equation}
where $\Phi_{\bmp}(\y_\ell)$ is either a Coulomb distorted wave
or simply the plane wave $e^{i{\bmp}\cdot{\y_\ell}}$ depending on whether
we are dealing with the $p+d$ or $n+d$ state ($\phi_{m_2}$ is the deuteron bound state wave function).
The final state consists of the lepton pair---with the $e^-$ having momentum (energy) ${\bm k}$ ($\epsilon$)
and spin $s$, and the $e^+$ having momentum (energy) ${\bm k}^\prime$ ($\epsilon^\prime$)
and spin $s^\prime$---and the trinucleon bound state recoiling with momentum $\bmp-\bmk-\bmk^\prime$.
We define hereafter $\bmq=\bmk+\bmk'$ and $\omega=\epsilon+\epsilon'$. Therefore,
$q^\mu q_\mu\equiv Q^2=\omega^2-q^2$. Energy conservation requires
\begin{equation}
\label{eq:e12}
\epsilon+\epsilon^\prime+ \frac{(\bmp-\bmk-\bmk^\prime)^2}{2\,M_3}= T_N+B_3-B_2\ ,
\end{equation}
where $M_3$ is the rest mass of the trinucleon ground state (either $\het$ or $\tri$)
and $B_3$ and $B_2$ are the binding energies of, respectively, the bound three-nucleon cluster and the deuteron.
Above, $T_N\equiv p^2/2m_N$ is the (laboratory) kinetic energy of the incident nucleon. Eq.~(\ref{eq:e12}) can be
rewritten as
\begin{eqnarray}
\epsilon+\epsilon^\prime&=&  E_0 -  \frac{q^2-2\bmp\cdot\bmq}{2\,M_3}\ ,\label{eq:e12b}\\
E_0&=&T_N\left(1-{m_N\over M_3}\right)+B_3-B_2\ .\label{eq:e12c}
\end{eqnarray}
Note that in good approximation $(1-m_N/M_3)=2/3$ and that $\epsilon+\epsilon^\prime\approx E_0$. 

The X17 is produced when $Q^2\approx M_X^2$. For this reaction $B_3-B_2\approx 5$ MeV, hence
the X17 could be produced only when $T_p\ge T_p^{min}$.
To find the value of $T_p^{min}$, it is convenient to consider the reaction in the
center-of-mass (CM) system.
At the threshold energy, the X17 and the trinucleon are produced at rest.
The energy conservation in this case reads
\begin{equation}
  T_{(CM)}^{min} -B_2 = M_X -B_3
\end{equation}
where $T_{(CM)}=(2/3)T_N$ is the initial CM kinetic energy of the $N+d$ pair.
Therefore, assuming for example $M_X=17$ MeV,
$T_{(CM)}^{min}=M_X-B_3+B_2=10.7$ MeV for $n+d$ and $11.5$ MeV for $p+d$. The minimum beam energies in the laboratory
system are therefore $16.1$ MeV for $n+d$ and $17.3$ MeV for $p+d$.

\begin{figure}[bth]
\centering
\includegraphics[scale=0.32,clip]{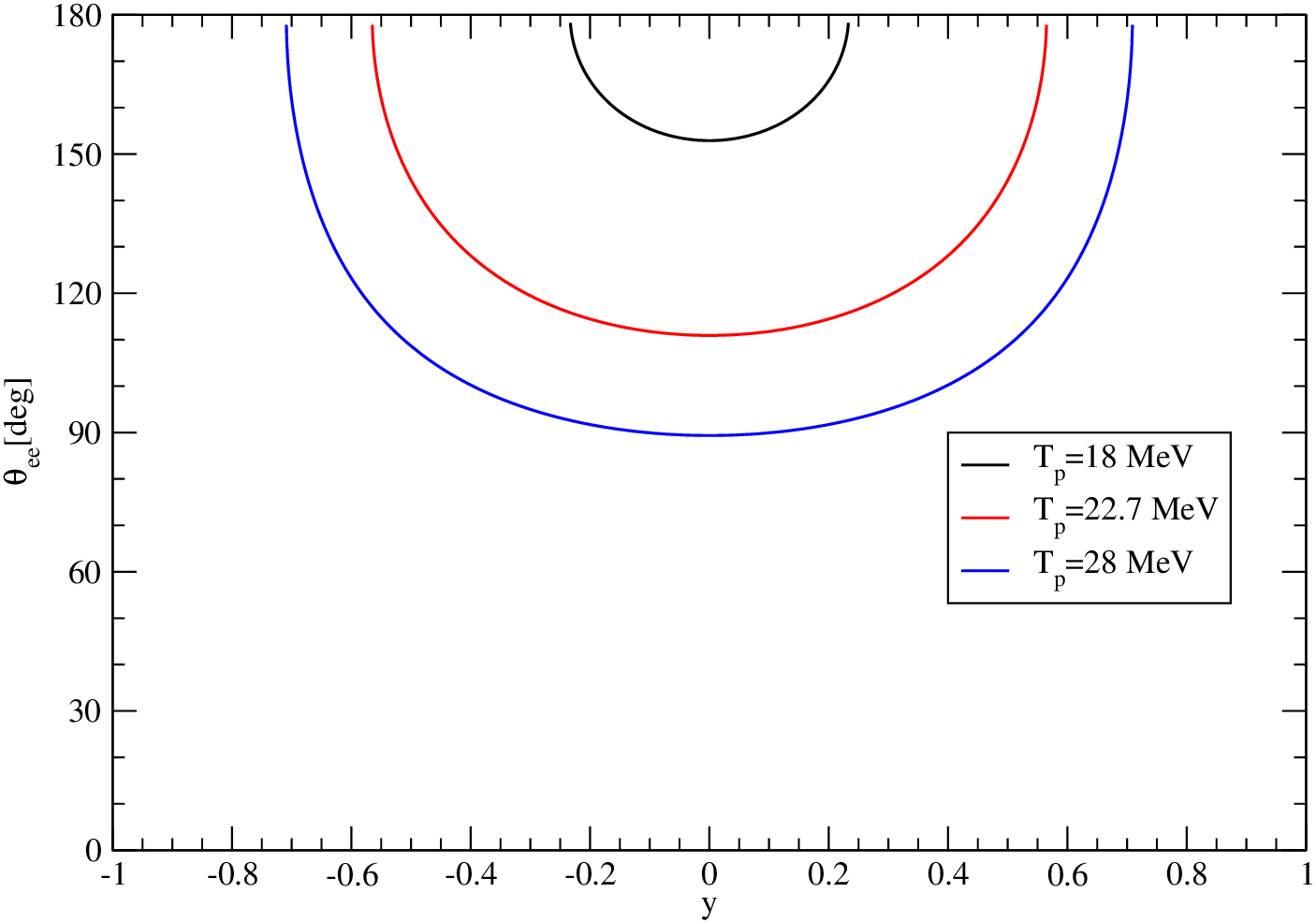}
\caption{(color online) The laboratory opening angle $\theta_{ee}$ between the two leptons in case
  of emission of an X17 as function
  of the quantity $y$, defined in Eq.~(\protect\ref{eq:y}), for different values of $E_0$.
  The relation between $E_0$ and $T_p$ is given in Eq.~(\protect\ref{eq:e12c}).
  Here we are considering the reaction $d(p,e^+e^-)\het$ and $M_X=17$ MeV.}
  \label{fig:ytheta}
\end{figure}

In case of production of an X17, the emission angle between the two leptons is practically determined by the
kinematics. For example, let us consider the emission of the two leptons in the plane perpendicular to
$\bmp$, the momentum of the incident beam, so that $\bmq\cdot\bmp=0$. Disregarding $q^2/2M_3$ in Eq.~(\ref{eq:e12b}), we have
$\omega=\epsilon+\epsilon^\prime\approx E_0$. 
As discussed above, the  X17 is produced when $Q^2\approx M_X^2$. This condition implies that
\begin{equation}
 \cos\theta_{ee} = {m_e^2 + {E_0^2\over 4} (1-y^2)-{M_X^2\over 2} \over
  \sqrt{{E_0^2\over4}(1+y)^2-m_e^2} \sqrt{{E_0^2\over4}(1-y)^2-m_e^2} } \ ,\label{eq:ytheta}
\end{equation}
where
\begin{equation}
  y={\epsilon-\epsilon^\prime\over \epsilon+\epsilon^\prime} \ . \label{eq:y}
\end{equation}
As the lepton energies vary between $m_e$ and $\approx E_0-m_e$, clearly $-1< y < +1$.
Eq.~(\ref{eq:ytheta}) fixes $\theta_{ee}$ in terms of $y$, $E_0$ and $M_X$.
For some values of these quantities, one may find that $|\cos\theta_{ee}|>1$. In such cases, the X17
emission is clearly kinematically forbidden. 
As an example, we show in Fig.~\ref{fig:ytheta} $\theta_{ee}$ vs $y$ for $M_X=17$ MeV and
for three different proton beam energies for the reaction $d(p,e^+e^-)\het$. As it can be seen, the minimum
opening angle between the two leptons (which corresponds to the peak in the angular distribution)
moves to lower and lower values as $E_0$ is increased.

\subsection{Decomposition of the matrix elements}

The matrix elements $j^\mu_{EM}$ and $j^c_X$ can be decomposed {\it i)} by expressing
the initial scattering wave function $\Psi_{m_2,m_1}$ in components
of definite total angular momentum and parity and {\it ii)} performing a
multipolar expansion of the operators (for more details, see Ref.~\cite{Viviani:2021stx}).
For the ``four-vector'' cases $j^\mu_{EM}$ and $j^{c=V,A}_{X}$, we consider the multipolar expansion 
of the charge, transverse, and longitudinal parts:
\begin{eqnarray}
\lefteqn{ \langle\Phi_{m_3}\,|\,\rho^\dag\,|\,\Psi_{m_2,m_1}\rangle =\qquad\qquad} &&  \nonumber \\
&&  \sum_{\ell m, LSJJ_z}4\pi(1,m_2,{1\over2},m_1| S,J_z)(L,0,S,J_z| J,J_z)\nonumber\\
&&\times    ({1\over2},m_3,J,-J_z| \ell,m) \sqrt{2L+1} i^L (-{i})^{\ell}  e^{i\sigma_L}\nonumber\\
&&\times   (-)^{J-J_z} e^{-im\phi_q} d_{-m,0}^{J}(-\theta_q) C_{\ell}^{LSJ}(q), \label{eq:c}\\
\lefteqn{ \langle\Phi_{m_3}\,|\, {\hat{\bme}}^{*}_{\lambda}\cdot \bmj^{\dag}(\bmq) \,|\,\Psi_{m_2,m_1}\rangle =\qquad\qquad} &&  \nonumber \\
&&  -\sum_{\ell m, LSJJ_z}\sqrt{8\pi^2} (1,m_2,{1\over2},m_1| S,J_z)(L,0,S,J_z| J,J_z)\nonumber\\
&&\times    ({1\over2},m_3,J,-J_z| \ell,m) \sqrt{2L+1} i^L (-{i})^{\ell}  e^{i\sigma_L}\nonumber\\
&&\times   (-)^{J-J_z} e^{-im\phi_q} d_{-m,-\lambda}^{J}(-\theta_q) \nonumber\\
&&\times \left[ \lambda \,M_{\ell}^{LSJ}(q) +E_{\ell}^{LSJ}(q)\right]\ ,
\label{eq:t}\\
\lefteqn{ \langle\Phi_{m_3}\,|\, {\hat{\bme}}^{*}_{z}\cdot \bmj^{\dag}(\bmq) \,|\,\Psi_{m_2,m_1}\rangle =\qquad\qquad} &&  \nonumber \\
&&  \sum_{\ell m, LSJJ_z}4\pi(1,m_2,{1\over2},m_1| S,J_z)(L,0,S,J_z| J,J_z)\nonumber\\
&&\times    ({1\over2},m_3,J,-J_z| \ell,m) \sqrt{2L+1} i^L (-{i})^{\ell}  e^{i\sigma_L}\nonumber\\
&&\times   (-)^{J-J_z} e^{-im\phi_q} d_{-m,0}^{J}(-\theta_q) L_{\ell}^{LSJ}(q) \ , \label{eq:l}
\end{eqnarray}
where $\lambda\,$=$\, \pm 1$, and $C_{\ell}^{LSJ}$,
$E_{\ell}^{LSJ}$, $M_{\ell}^{LSJ}$, and $L^{LSJ}_\ell$ denote the reduced matrix 
elements (RMEs) of the charge $(C)$, transverse electric $(E)$,
transverse magnetic $(M)$, and longitudinal $(L)$ multipole operators, defined as in
Ref.~\cite{Walecka1995}. Above, $\sigma_L$ is the Coulomb phase shift and we have introduced the basis of unit vectors
\begin{equation}
\hat{\bm e}_{z}=\hat\bmq\ ,\qquad \hat{\bm e}_{y}=\frac{{\bm p}\times\bmq}{|{\bm p}\times\bmq|}
\ ,\qquad \hat{\bm e}_{x}=\hat{\bm e}_{y}\times\hat{\bm e}_{z} \ ,
\end{equation}
and ${\bm e}_\pm=\mp (\hat{\bm e}_x\pm i\, \hat{\bm e}_y)/\sqrt{2}$.
Clearly, for the cases $c=S$ and $P$, only the charge matrix elements of Eq.~(\ref{eq:c}) are needed.
Moreover, in the EM case, using current conservation, it is possible to write
the matrix elements of the longitudinal component of the current operator in terms of those of the charge operator. 

In the matrix elements above, the spin quantization axis of the nuclear states is taken along
the incident nucleon momentum ${\bm p}\,$=$\,p\, \hat{\bm z}$ rather
than the three-momentum transfer $\bmq\,$=$\,q\, \hat{\bm e}_z$, as usual.
For this reason, we needed to introduce Wigner rotation 
matrices ${d}_{M',M}^{J}$~\cite{Edmond1957}. The angles
$\theta_q$ and $\phi_q$ specify the direction of $\bmq$
in the lab frame (with ${\bm p}$ along $\hat{\bm z}$).
Finally, the RMEs are computed in a frame where $\bmq$ is along $z$, where
$d^J_{M',M}(0)=\delta_{M',M}$.

\subsection{The cross section for $d(p,e^+e^-)\het$ and $d(n,e^+e^-)\tri$ processes}
The general expression of the five-fold differential cross section can be schematically written as
\begin{eqnarray}
  \frac{d^5\sigma}{d\epsilon\,d\hat{\bmk}\, d\hat{\bmk}^\prime}&=&
    {2\over 3(2\pi)^3} {\alpha^2\over v} kk' f_{\rm rec} \Bigl[
      {R_{EM}(\epsilon,\hat{\bmk}, \hat{\bmk}^\prime) \over Q^4}\nonumber\\
      &+&  
  {\varepsilon_e^c R_X(\epsilon,\hat{\bmk}, \hat{\bmk}^\prime)\over Q^2 D_X}+c.c.
  +{(\varepsilon_e^c)^2 R_{XX}(\epsilon,\hat{\bmk}, \hat{\bmk}^\prime)\over |D_X|^2}\Bigr] \nonumber\\
    &=&
    {2\over 3(2\pi)^3} {\alpha^2\over v} kk' f_{\rm rec} \Bigl[
      {R_{EM}(\epsilon,\hat{\bmk}, \hat{\bmk}^\prime)\over Q^4}+\nonumber\\
 &+&        
      {\varepsilon_e^c R_X(\epsilon,\hat{\bmk}, \hat{\bmk}^\prime)D_X^*/Q^2+c.c.\over |D_X|^2}\nonumber\\
 &+&
      {(\varepsilon_e^c)^2  R_{XX}(\epsilon,\hat{\bmk}, \hat{\bmk}^\prime)\over |D_X|^2}\Bigr]
    \ ,  \label{eq:e13}
\end{eqnarray}
where $v$ is the $N+d$ relative velocity, $D_X=Q^2-M_X^2$ (we remember that $Q^2=q^\mu q_\mu$) and the three terms
denote the contributions coming solely from EM currents,
the interference between EM and X17-induced currents, and
purely X17-induced currents, respectively. The positron
energy $\epsilon^\prime$ is fixed by energy conservation. In the 
laboratory coordinate system, where the $z$-axis is oriented
along the incident beam momentum $\bmp$, the spherical angles 
specifying the $\hat\bmk$ ($\hat\bmk'$) direction are denoted
as $\theta$ and $\phi$ ($\theta'$ and $\phi'$), and
\begin{equation}
 \hat\bmk \cdot \hat\bmk'  \equiv \cos\theta_{ee}=\cos\theta\cos\theta'+\sin\theta\sin\theta'\cos(\phi'-\phi)\ .
    \label{eq:thetaee}
\end{equation}
In Eq.~(\ref{eq:e13}) we have made explicit the dependence on the X17-electron coupling constant $\varepsilon_e^c$.
The quantity $R_{X}$ ($R_{XX}$) depends linearly (quadratically) on the X17-hadron coupling constants $\eta^c_{0,z}$.
Finally, the recoil factor is 
\begin{equation}
  f_{\rm rec}^{-1}=\left | 1 +{1\over M_3}(k'-p \cos\theta' +k \cos\theta_{ee}) {\epsilon'\over k'}\right|\ ,
\end{equation}
where $\theta'$ is the angle between the directions of the positron and incident nucleon momenta, and $\theta_{ee}$
is the angle between the momenta of the two leptons defined in Eq.~(\ref{eq:thetaee}).

The quantities $R_{EM}$, $R_{X}$, and $R_{XX}$ can be obtained as combinations of the
matrix elements given in Eqs.~(\ref{eq:c})--(\ref{eq:l}) and kinematical terms depending on the leptonic variables. For example,
\begin{equation}
R_{EM}=\sum_{n=1}^6 v_n\, R_n \ ,
\end{equation}
where
\begin{eqnarray}
\label{eq:e30}
  v_1 &=& (Q^4/q^4)(\epsilon\epsilon^\prime+ \bmk\cdot\bmk^\prime-m_e^2)\ , \nonumber\\
  v_2&=& -P_x\,[ \epsilon-\epsilon^\prime-(\omega/q)\, P_z ]/\sqrt{2} \ , \nonumber\\
  v_3&=& -P_y\,[ \epsilon-\epsilon'-(\omega/q) \,P_z]/\sqrt{2} \ ,\label{eq:vi}\\
  v_4&=& -(P_x^2+P_y^2)/4 + m_e^2 + \epsilon\epsilon' - \bmk\cdot\bmk'\ , \nonumber\\
  v_5&=& (P_x^2-P_y^2)/2\ , \nonumber\\
  v_6&=& -P_x P_y \ , \nonumber
\end{eqnarray}
and
\begin{eqnarray}
\label{eq:e31}
  R_1 &=& \sum_{m_3,m_1} |\rho_{fi}|^2\ , \nonumber\\
  R_2 &=& \sum_{m_3,m_1} {\rm Re}\, [ \,\rho_{fi}^*\,  (\, j^+_{fi}-j^-_{fi}\,) \,]\ , \nonumber\\
  R_3 &=& \sum_{m_3,m_1} {\rm Im}\,[ \,\rho_{fi}^*\,  (\, j^+_{fi}+j^-_{fi}\,) \,]\ ,\label{eq:Ri}\\
  R_4 &=& \sum_{m_3,m_1} (\, |j^+_{fi}|^2 + |j^-_{fi}|^2 \,)\ ,\nonumber \\
  R_5 &=& \sum_{m_3,m_1} {\rm Re}\, (\,j_{fi}^{+\,*} \,j^-_{fi}\, )\ , \nonumber\\
  R_6 &=& \sum_{m_3,m_1} {\rm Im}\,(\,j_{fi}^{+\,*} \,j^-_{fi}\, ) \ .\nonumber
\end{eqnarray}
Above $\rho_{fi}$ and $j^\pm_{fi}$ are the matrix elements given in Eqs.~(\ref{eq:c}) and~(\ref{eq:t}), respectively,
while $P_a$ denotes the component of
${\bm P}={\bmk}-{\bmk}^\prime$ along ${\bm e}_a$. The expressions for
$R_{X}$ and $R_{XX}$ are explicitly given in Ref.~\cite{Viviani:2021stx}.

The four-fold differential cross section is obtained
by integrating over the electron energy,
\begin{equation}
  {d^4\sigma\over d\hat \bmk \,d\hat\bmk'}= 
  \int_{m_e}^{\epsilon_{max}} d\epsilon\,   \frac{d^5\sigma}{d\epsilon\,d\hat{\bmk}\, d\hat{\bmk}^\prime}
  \ ,\label{eq:xs4}
\end{equation}
where the maximum allowed energy $\epsilon_{max}$ is obtained from the
solution of Eq.~(\ref{eq:e12}) for the case of a vanishing positron
momentum; in fact, since the kinetic energy of the recoiling trinucleon bound state
is tiny, $\epsilon_{max}$ is close to $E_0-m_e$. 

To account for the decay of X17, we introduce a width $\Gamma_X$ and will make in Eq.~(\ref{eq:e13}) the replacement
\begin{equation}
   \label{eq:e3}
   M_X\longrightarrow M_X-i\, \Gamma_X/2 \ .
\end{equation}
Hence, $ D_X= Q^2-M_X^2 +{\Gamma_X^2\over4} +i M_X \Gamma_X$.
Assuming that the predominant decay channel of X17 is in $e^+e^-$~\cite{Feng:2016jff}, then
\begin{equation}
  \Gamma_X\sim \alpha\,(\varepsilon_e^c)^2 \, M_X\ .\label{eq:gammaX}
\end{equation}
Given the current bounds on $\varepsilon_e^c$ ($\sim 10^{-3}$ as discussed in Sec.~\ref{sec:intro}),
we have $\Gamma_X\ll M_X$ and $D_X\approx Q^2-M_X^2+i M_X \Gamma_X$. Now, the X17 contribution to the cross section
(the two terms $\sim 1/|D_X|^2$) is only sizeable where $Q^2-M_X^2=0$~\cite{Viviani:2021stx}. We can distinguish two
kinematical regions of $\epsilon$, $\hat{\bmk}$, and $\hat{\bmk}^\prime$ values:
\begin{itemize}
\item \underline{Region A}, where the $Q^2-M_X^2=0$ is never satisfied;
  in that region, the contribution of the X17 is always negligible with respect to the EM  one, therefore
    here we can safely assume that
   \begin{equation}
     \frac{d^5\sigma}{d\epsilon\,d\hat{\bmk}\, d\hat{\bmk}^\prime}=
    {2\over 3(2\pi)^3} {\alpha^2\over v} kk' f_{\rm rec} 
      {R_{EM}(\epsilon,\hat{\bmk}, \hat{\bmk}^\prime) \over Q^4}
     \ .\label{eq:regA}
    \end{equation}
  \item  \underline{Region B}, where $Q^2-M_X^2=0$ is satisfied for some values of $\epsilon$, $\hat{\bmk}$, and $\hat{\bmk}^\prime$.
  For fixed values of $\hat{\bmk}$, and $\hat{\bmk}^\prime$, the condition $Q^2-M_X^2=0$
  is verified for two values of $\epsilon$, denoted in the following as $\epsilon_i$, $i=1,2$~\cite{Viviani:2021stx};
  since $\Gamma_X$ is very small, $1/|D_X|^2$ is always negligible except for $\epsilon\approx\epsilon_i$, where
  it assumes the form of a very narrow Lorentzian, which can be very well represented by a delta function; namely, 
 \begin{equation}
   \frac{1}{|D_X|^2} \longrightarrow \sum_{i=1,2}{\gamma_i\over (\varepsilon_e^c)^2}\delta(\epsilon-\epsilon_i) \ ,
 \label{eq:eeee}
 \end{equation}
 where $\gamma_i$ are factors independent on $\epsilon$ and $\varepsilon_e^c$~\cite{Viviani:2021stx}
 (clearly, $\epsilon_i$ and $\gamma_i$
 depend on the given choice of $\hat{\bmk}, \hat{\bmk}^\prime$).
 Above, we have taken into account the dependence of $\Gamma_X$ on $\varepsilon_e^c$ given in
 Eq.~(\ref{eq:gammaX}).
 It is worthwhile to point out that, in region B, the interference
  contribution (proportional to $R_X$) is always negligible relative to the $R_{XX}$ term,
  as discussed in detail in Appendix~\ref{app:b}. Therefore, we can write
   \begin{eqnarray}
       \frac{d^5\sigma}{d\epsilon\,d\hat{\bmk}\, d\hat{\bmk}^\prime}&=&
    {2\over 3(2\pi)^3} {\alpha^2\over v} kk' f_{\rm rec} \Bigl[
      {R_{EM}(\epsilon,\hat{\bmk}, \hat{\bmk}^\prime) \over Q^4}\nonumber\\
    &&+ \sum_{i=1,2} R_{XX}(\epsilon_i,\hat{\bmk}, \hat{\bmk}^\prime)\gamma_i
     \delta(\epsilon-\epsilon_i)\Bigr]\ . \label{eq:regB}
   \end{eqnarray}
\end{itemize}
Note that, in the present tree-level treatment of the X17 width, the cross-section becomes independent
on $\varepsilon_e^c$. 

In order to obtain the four-fold differential cross section, the integration
over $\epsilon$ is carried out numerically for the EM
term, and analytically for the term with $R_{XX}$ (if present).
In next section, we will mostly present the calculated  four-fold differential cross sections for
the leptons emitted in the perpendicular plane with respect to
the incident nucleon momentum (i.e. for $\theta=\theta'=90$ deg) and vs.\ $\theta_{ee}$.
In some cases, we also present the results for other values of $\theta$ and $\theta'$.
We account roughly for a (possible) finite angular resolution of the detector
employed in an eventual future experiment by folding the $\theta_{ee}$ dependence
of this (four-fold) cross section with a normalized Gaussian of width $\Delta$,
chosen to be $\Delta=2.5$ deg. This folding has practically no effect on
the EM part, as the dependence on $\theta_{ee}$ in this case is rather flat,
but it is effective to smooth out the peak due to the exchange of the X17, 
see Ref.~\cite{Viviani:2021stx} for more details. 

Finally, we will report also the results for the total cross sections of the $d(p,\gamma)\het$
and $d(n,\gamma)\tri$ radiative captures. They are given by
\begin{equation}
  \sigma_C \!=\! {16\pi^2\alpha\over 3v}  {q\over 1+q/M_3} \!\!\sum_{\ell LS,J\geq1}
  \Bigl[ |E_\ell^{LSJ}(q)|^2 + |M_\ell^{LSJ}(q)|^2\Bigr] ,\label{eq:capture}
\end{equation}
where $q$ is the momentum of the outgoing photon and the sum only includes
EM transverse RMEs.

\section{Results}
\label{sec:res}

We list in Table~\ref{tab:rmes} the RMEs contributing to the transition
from an initial $^{2S+1}L_J$ $2+1$ scattering state to the final trinucleon ground state
with $J^\pi\,$=$\,{1\over2}^+$. Note that the multipolarity $\ell$ of the RME
has to be in the range $|J-{1\over2}|\le \ell \le J+{1\over2}$.

\begin{table}[bth]
  \caption{\label{tab:rmes}
   The RMEs $C^{LSJ}_\ell$, $E^{LSJ}_\ell$, $M^{LSJ}_\ell$, and $L^{LSJ}_\ell$ contributing
   to the EM and X17 transitions
   from an initial $2+1$ $^{2S+1}L_J$ scattering state to the final
   trinucleon ground state.}
    \begin{center}
      \begin{tabular}{l|c|cccc}
        \hline\hline
        \multicolumn{6}{c}{EM,S,V cases}\\
        \hline\hline
        state & ${}^{2S+1}L_J$ & $C_\ell^{LSJ}$ & $E_\ell^{LSJ}$ & $M_\ell^{LSJ}$ & $L_\ell^{LSJ}$ \\
        \hline
        ${1\over2}^+$  &  ${}^2S_{1\over2}, {}^4D_{1\over2}$ & $C_0^{LS{1\over2}}$ & $-$ & $M_1^{LS{1\over2}}$ & $L_0^{LS{1\over2}}$\\
        ${1\over2}^-$  &  ${}^2P_{1\over2}, {}^4P_{1\over2}$ & $C_1^{LS{1\over2}}$ & $E_1^{LS{1\over2}}$ & $-$ & $L_1^{LS{1\over2}}$ \\
        \hline
        ${3\over2}^+$  &  ${}^4S_{3\over2}, {}^2D_{3\over2}, {}^4D_{3\over2}$ & $C_2^{LS{3\over2}}$ & $M_1^{LS{3\over2}}$
                                                                              & $E_2^{LS{3\over2}}$ & $L_2^{LS{3\over2}}$\\
        ${3\over2}^-$  &  ${}^2P_{3\over2}, {}^4P_{3\over2}, {}^4F_{3\over2}$ & $C_1^{LS{3\over2}}$ & $M_2^{LS{3\over2}}$
                                                                              & $E_1^{LS{3\over2}}$ & $L_1^{LS{3\over2}}$\\
        \hline
        ${5\over2}^+$  &  ${}^2D_{5\over2}, {}^4D_{5\over2}, {}^4G_{5\over2}$ & $C_2^{LS{5\over2}}$ & $M_3^{LS{5\over2}}$
                                                                              & $E_2^{LS{5\over2}}$ & $L_2^{LS{5\over2}}$\\
        ${5\over2}^-$  &  ${}^4P_{5\over2}, {}^2F_{5\over2}, {}^4F_{5\over2}$ & $C_3^{LS{5\over2}}$ & $M_2^{LS{5\over2}}$
                                                                              & $E_3^{LS{5\over2}}$ & $L_3^{LS{5\over2}}$\\
        \hline\hline
        \multicolumn{6}{c}{P,A cases}\\
        \hline\hline
        state & ${}^{2S+1}L_J$ & $C_\ell^{LSJ}$ & $E_\ell^{LSJ}$ & $M_\ell^{LSJ}$ & $L_\ell^{LSJ}$ \\
        \hline
        ${1\over2}^+$  &  ${}^2S_{1\over2}, {}^4D_{1\over2}$ & $C_1^{LS{1\over2}}$ & $E_1^{LS{1\over2}}$ & $-$ & $L_1^{LS{1\over2}}$\\
        ${1\over2}^-$  &  ${}^2P_{1\over2}, {}^4P_{1\over2}$ & $C_0^{LS{1\over2}}$ & $-$ & $M_1^{LS{1\over2}}$ & $L_0^{LS{1\over2}}$ \\
        \hline
        ${3\over2}^+$  &  ${}^4S_{3\over2}, {}^2D_{3\over2}, {}^4D_{3\over2}$ & $C_1^{LS{3\over2}}$ & $M_2^{LS{3\over2}}$
                                                                              & $E_1^{LS{3\over2}}$ & $L_1^{LS{3\over2}}$\\
        ${3\over2}^-$  &  ${}^2P_{3\over2}, {}^4P_{3\over2}, {}^4F_{3\over2}$ & $C_2^{LS{3\over2}}$ & $M_1^{LS{3\over2}}$
                                                                              & $E_2^{LS{3\over2}}$ & $L_2^{LS{3\over2}}$\\
        \hline
        ${5\over2}^+$  &  ${}^2D_{5\over2}, {}^4D_{5\over2}, {}^4G_{5\over2}$ & $C_3^{LS{5\over2}}$ & $M_2^{LS{5\over2}}$
                                                                              & $E_3^{LS{5\over2}}$ & $L_3^{LS{5\over2}}$\\
        ${5\over2}^-$  &  ${}^4P_{5\over2}, {}^2F_{5\over2}, {}^4F_{5\over2}$ & $C_2^{LS{5\over2}}$ & $M_3^{LS{5\over2}}$
                                                                              & $E_2^{LS{5\over2}}$ & $L_2^{LS{5\over2}}$\\
        \hline
      \end{tabular}    
    \end{center}
  \end{table}

\subsection{Results for the electromagnetic IPC}
\label{sec:res_em_rmeaa}

In the EM case, the long-wavelength approximation (of relevance here) relates the electric and Coulomb 
operators, respectively $\hat E_{JM}(q)$ and $\hat C_{JM}(q)$, via~\cite{Walecka1995}
\begin{equation}
  \hat E_{JM} (q) \approx \sqrt{J+1\over J} {\Delta E\over q}\, \hat C_{JM}(q)\ ,
  \label{eq:siegert}
\end{equation}
where $\Delta E=E_i-E_f$ is the difference between the initial $2+1$ scattering state
and trinucleon ground state energies (Siegert's
theorem~\cite{Siegert1937}). This relation implies a relationship between
the corresponding EM RMEs $E^{LSJ}_\ell(q)$ and $C^{LSJ}_\ell(q)$. It is worthwhile stressing here
that Siegert's theorem assumes (i) a conserved current and (ii) that
the initial and final states are exact eigenstates of the nuclear
Hamiltonian.  Eq.~(\ref{eq:siegert}) provides a
test---indeed, a rather stringent one---of these two assumptions, see Ref.~\cite{Schiavilla:2018udt}
for a discussion of this issue in the context of the chiral interaction
NVIa and accompanying EM currents. 
  
\begin{figure}[bth]
\centering
\includegraphics[scale=0.40,clip]{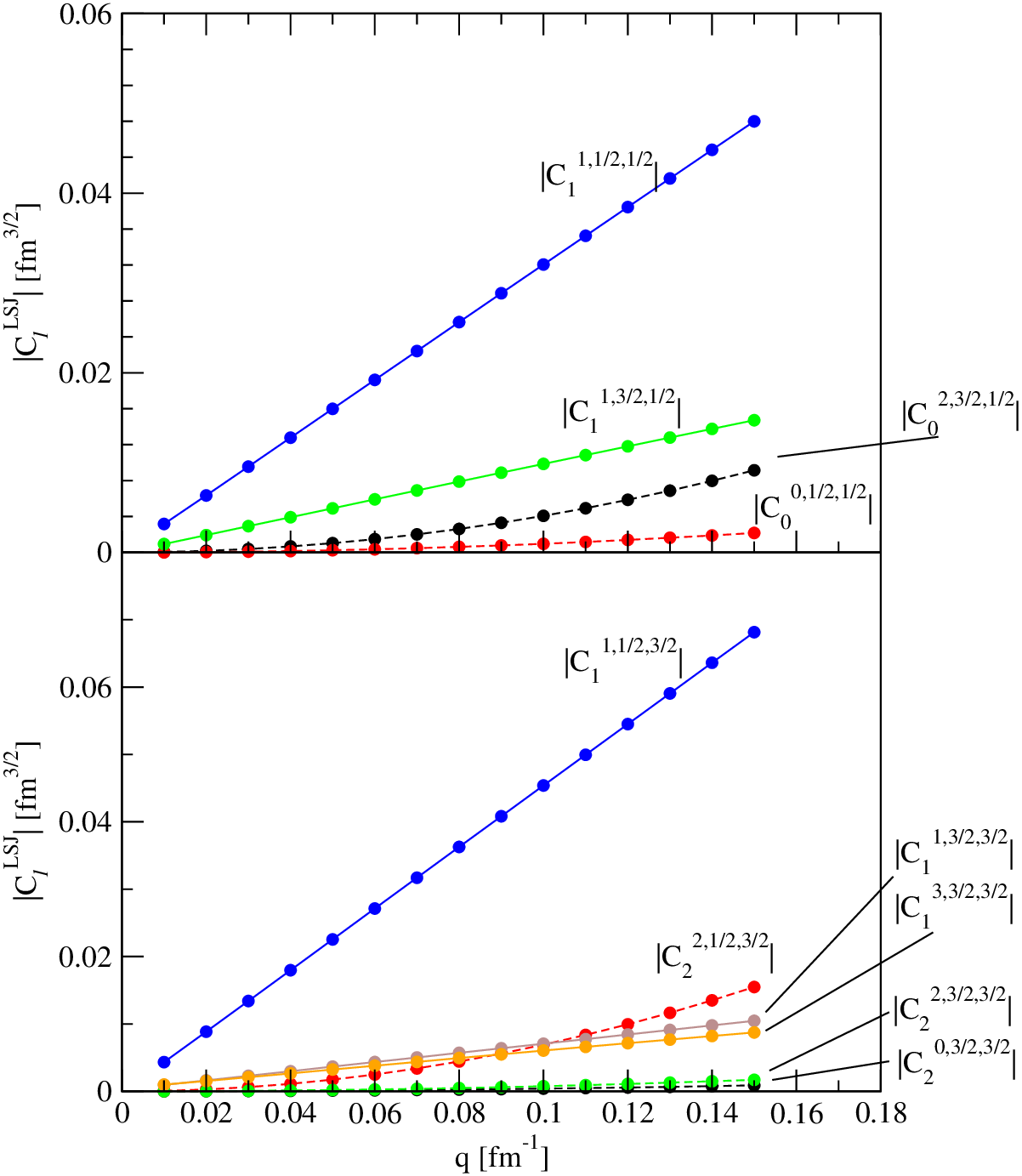}
\caption{(color online) The dependence on the three-momentum transfer $q$ of some charge EM RMEs (solid circles);
  the calculations are at incident proton energy of 18 MeV and use the AV18UIX interaction.
  The solid (dashed) lines show fits of the calculated values using linear (quadratic) parametrizations.
  Smaller RMEs are not shown.}
  \label{fig:qdep1}
\end{figure}

\begin{figure}[bth]
\centering
\includegraphics[scale=0.40,clip]{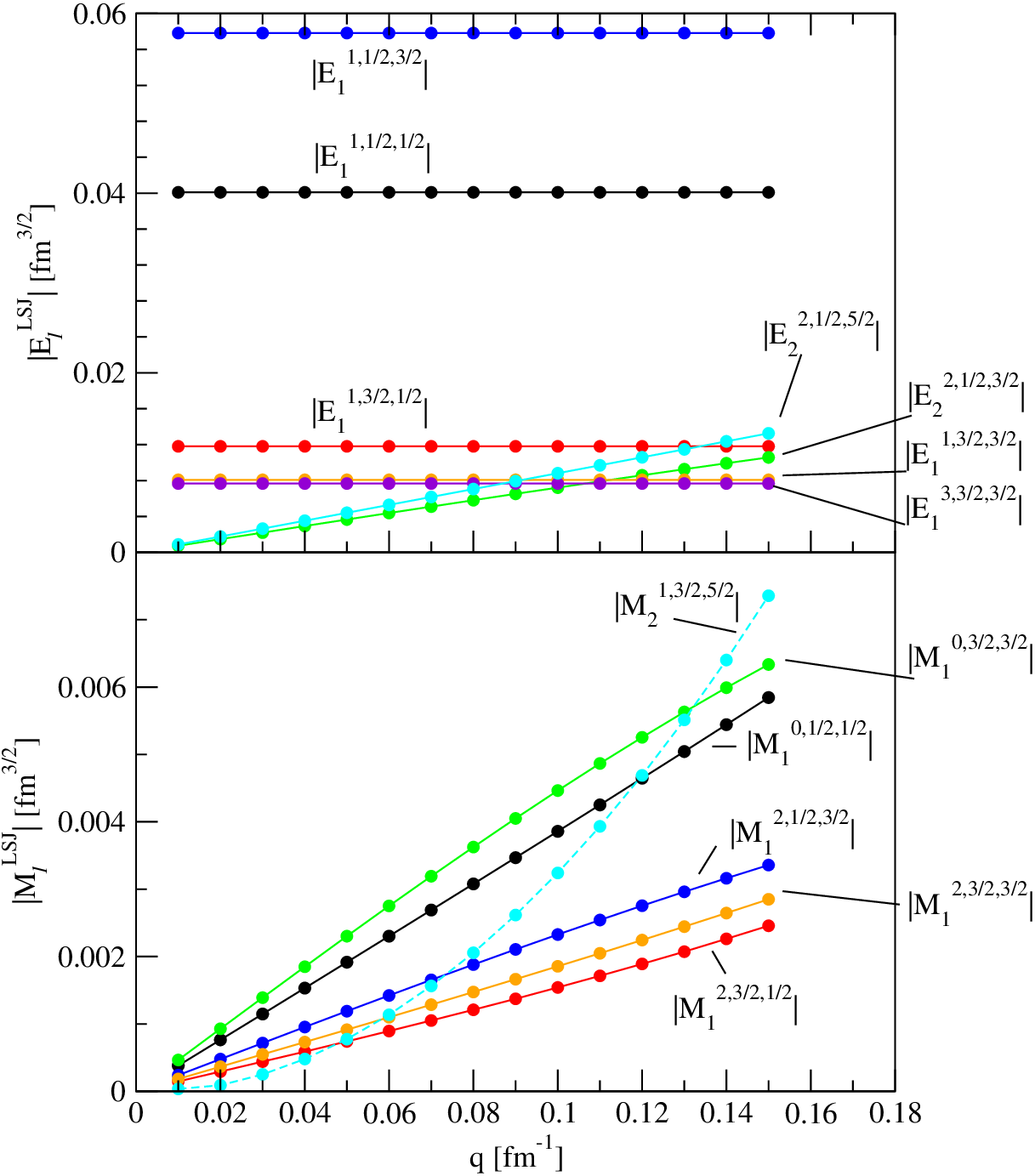}
\caption{(color online) The same as in Fig.\protect\ref{fig:qdep1} but for the electric and
magnetic EM RMEs.}
  \label{fig:qdep2}
\end{figure}

We report in Figs.~\ref{fig:qdep1} and~\ref{fig:qdep2} the EM RMEs calculated with the AV18UIX
interaction for $p+d$ at $T_p=18$ MeV.
From the figures, we can see that the largest RMEs are the $C_1$'s and $E_1$'s coming from the
transitions  ${1\over2}^-\longrightarrow {1\over2}^+$ and ${3\over2}^-\longrightarrow{1\over2}^+$.
In fact, in this range of energies we are in the region of the giant dipole resonance,
hence the importance of the $E_1$ transitions. To understand the $q$ dependence of this transition, we
recall that $C_1$ RME involves the matrix element of 
the operator proportional to $\sum_i {1\over2}\Bigl(1+\tau_z(i)\Bigr) j_1(qr_i)$, where $j_1$ is a spherical Bessel function
of order 1. Therefore, at small $q$, we deduce that these $C_1\sim q$. From Eq.~(\ref{eq:siegert}), then we obtain that $E_1\sim q^0$.
We note also that $|E_1^{1,{1\over2},J}|> |E_1^{1,{3\over2},J}|$ for both $J={1\over2}$ and ${3\over2}$.
This is related to the fact that the $E_1$ operator is essentially spin-independent. As a consequence,
such an operator can connect the large $S$-wave component having
total spin $S$=${1\over2}$ in the trinucleon ground states to the $^2P_J$ scattering state.
However, this does not happen for the $^4P_J$ scattering state because of orthogonality between the spin states.
Consequently, the transitions from the $^4P_J$ scattering states proceed only via the small components
of the trinucleon ground state (these components account for roughly 8\% of the trinucleon normalizations).

The transition ${1\over2}^+\longrightarrow {1\over2}^+$ is suppressed for two reasons. At these energies, the
$N-d$ interaction in this wave is repulsive or only slightly attractive. Moreover,
at LO $C_0^{L,S,{1\over2}}$ involves the matrix
element of the  $\hat C_0$ multipole operator proportional to
$\sum_i {1\over 2}\Bigl(1+\tau_z(i)\Bigr) j_0 (qr_i)$ between the $N+d$ scattering and trinucleon ground state
wave functions. In the $q$ expansion of the spherical Bessel function, the ``leading'' term is
$\sum_i {1\over 2}\Bigl(1+\tau_z(i)\Bigr)={3\over2}+\hat T_z $, where $\hat T_z$ is the $z$-component of the
total isospin operator. Since $\hat T_z|N+d\rangle=\pm {1\over 2}$, then this leading term gives a
vanishing contribution to the matrix element because of the orthogonality between the ground and scattering states,
and hence the $C_0^{L,S,{1\over2}}$ RME is proportional to $q^2$. We note that at these
energies, it is even found that $|C_0^{2,{3\over2},{1\over2}}|> |C_0^{0,{1\over2},{1\over2}}|$.

Regarding the transition  ${3\over2}^+\longrightarrow {1\over2}^+$, the contribution of the  ${}^4S_{3\over2}$ wave
is suppressed since the Pauli principle forbids identical nucleons
with parallel spins to come close to each other. In fact, the RME
$C_2^{0,{3\over2},{3\over2}}$ is rather suppressed even with respect to
$C_2^{2,{1\over2},{3\over2}}$.

Higher-order transitions with $\ell\ge 2$ are usually suppressed
by powers of the three-momentum transfer $q$ which is $\lesssim 0.1$ fm$^{-1}$,
the only exception being the $E_2$ transition involving the $^2D_{5\over2}$ channel,
this case being favoured by the fact that the Pauli principle does not play any role, and
the scattering state being a $S={1\over2}$ state. 
We note finally that all the magnetic RMEs result to be rather small. 

\begin{figure}[t]
\centering
\includegraphics[scale=0.35,clip]{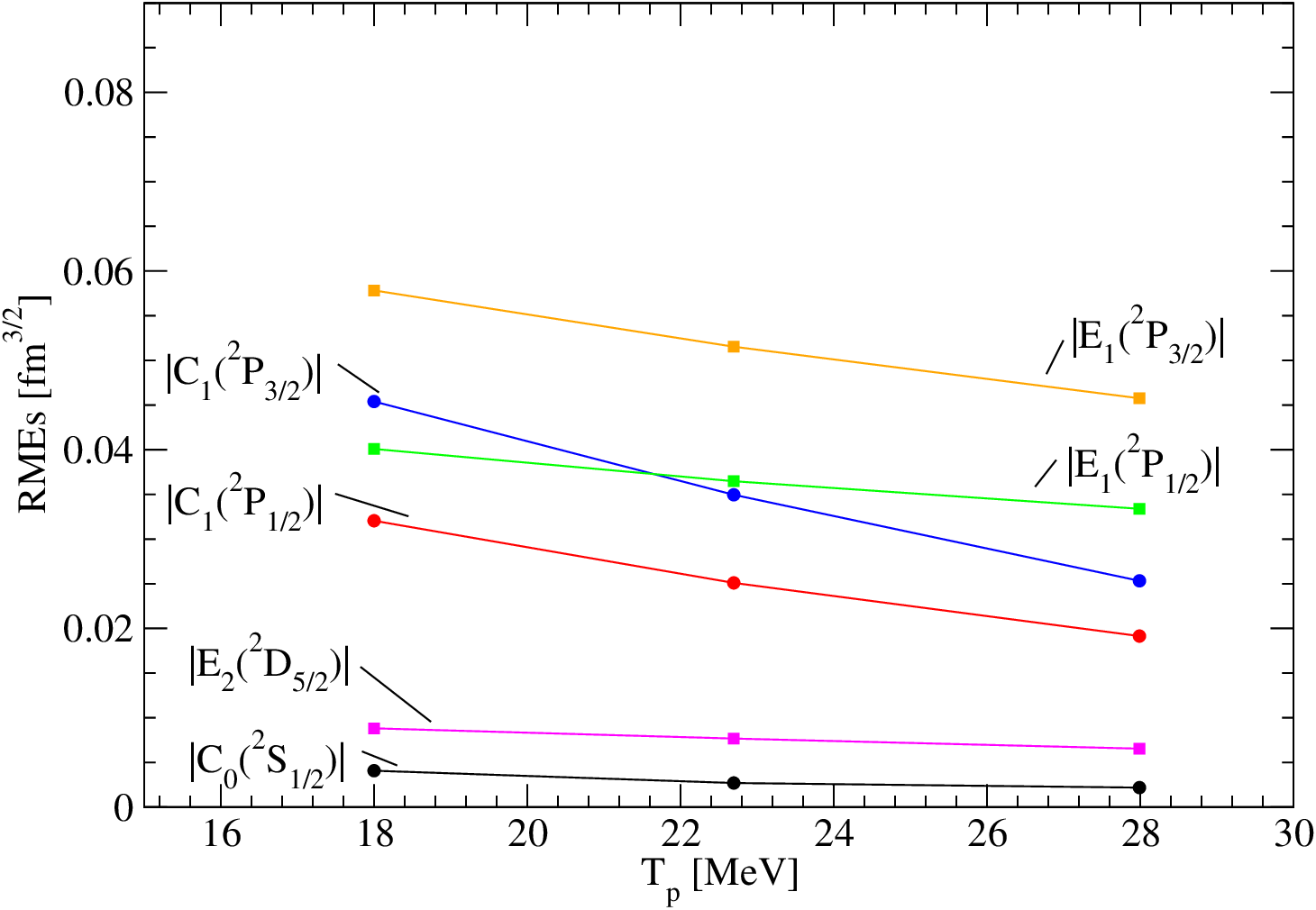}
\caption{(color online) The dependence on the proton incident energy $T_p$ of some EM RMEs (solid symbols); the calculations
are at a fixed three-momentum transfer $q=0.1$ fm$^{-1}$ and use 
the AV18UIX interaction.  The lines are to guide the eyes only.}
  \label{fig:Edep}
\end{figure}

Figure~\ref{fig:Edep} shows the behavior of selected $p+d$ EM RMEs (the largest ones) as function
of the incident proton energy.  As it can be seen, the
behaviour of these (and also of the others) RMEs is
monotonically decreasing, reflecting the absence of resonances in the trinucleon spectrum.

Let us compare  now the $p+d$ and $n+d$ RMEs. In Table~\ref{tab:pdnd}, we report a selected set
of RMEs calculated for both reactions for a beam energy of 18 MeV. As it can be seen, the largest
$E_1$ RMEs are almost identical (due to the fact that the $C_1$ are related to the $E_1$, the same is
observed for these RMEs). This can be understood since the $E_1$ operator is essentially an isovector,
therefore $E_1^{LSJ}(p+d)\approx -E_1^{LSJ}(n+d)$. For other RMEs, usually the $p+d$ are larger 
(often approximately by a factor 2) than those for $n+d$, as one could naively expect since the $p+d$
reactions involves two protons. However, since the $E_1$ and $C_1$ RMEs coming from the transitions
${1\over2}^-\longrightarrow{1\over2}^+$ and ${3\over2}^-\longrightarrow{1\over2}^+$ are dominant,
from this result one can expect that the IPC cross section be approximately the same. 

\begin{table}[bth]
  \caption{\label{tab:pdnd}
A selected set of RMEs in absolute value (in fm$^{3/2}$) corresponding to the $p+d$ and $n+d$ reactions,
obtained with the AV18UIX and NVIa3N interactions and accompanying EM charge and current operators.
The incident nucleon energy is 18 MeV and the three-momentum transfer $q$ is 0.1 fm$^{-1}$.}  
  \begin{center}
    \begin{tabular}{l|c|cc|cc}
            \hline
            \hline
           & & \multicolumn{2}{c}{AV18UIX} & \multicolumn{2}{c}{NVIa3N} \\
      RMEs$\times 10^3$ & wave & $p+d$ &  $n+d$ & $p+d$ &  $n+d$ \\      
      \hline
      $|C_0^{0,{1\over2},{1\over2}}|$ & ${}^2S_{1\over2}$ &  $ 4.11$ & $ 2.12$ & $ 4.12$ & $ 2.16$ \\
      $|C_0^{2,{3\over2},{1\over2}}|$ & ${}^2S_{1\over2}$ &  $ 0.97$ & $ 0.30$ & $ 0.99$ & $ 0.30$ \\
      \hline
      $|E_1^{1,{1\over2},{1\over2}}|$ & ${}^2P_{1\over2}$ &  $39.36$ & $40.20$ & $40.42$ & $40.32$ \\
      $|E_1^{1,{3\over2},{1\over2}}|$ & ${}^4P_{1\over2}$ &  $12.07$ & $12.45$ & $12.00$ & $12.48$ \\
      \hline
      $|E_1^{1,{1\over2},{3\over2}}|$ & ${}^2P_{3\over2}$ &  $57.43$ & $57.96$ & $58.36$ & $58.07$ \\
      $|E_1^{1,{3\over2},{3\over2}}|$ & ${}^4P_{3\over2}$ &  $ 8.02$ & $ 8.66$ & $ 8.04$ & $ 8.64$ \\
      \hline
      $|E_2^{2,{1\over2},{5\over2}}|$ & ${}^2D_{5\over2}$ &  $ 8.79$ & $ 1.92$ & $ 8.81$ & $ 1.91$ \\
      $|M_2^{1,{3\over2},{5\over2}}|$ & ${}^4P_{5\over2}$ &  $ 3.20$ & $ 2.18$ & $ 3.25$ & $ 2.18$ \\
      \hline
      \hline
    \end{tabular}
    \end{center}
  \end{table}

%
%

Finally, we calculate the cross-sections obtained for the
internal pair conversion processes.  The calculations use fully converged
bound- and scattering-state wave functions (with the largest allowed
number of HH states) and the complete N4LO set of EM
charge and current operators.

\begin{figure}[bth]
\centering
\includegraphics[scale=0.35,clip]{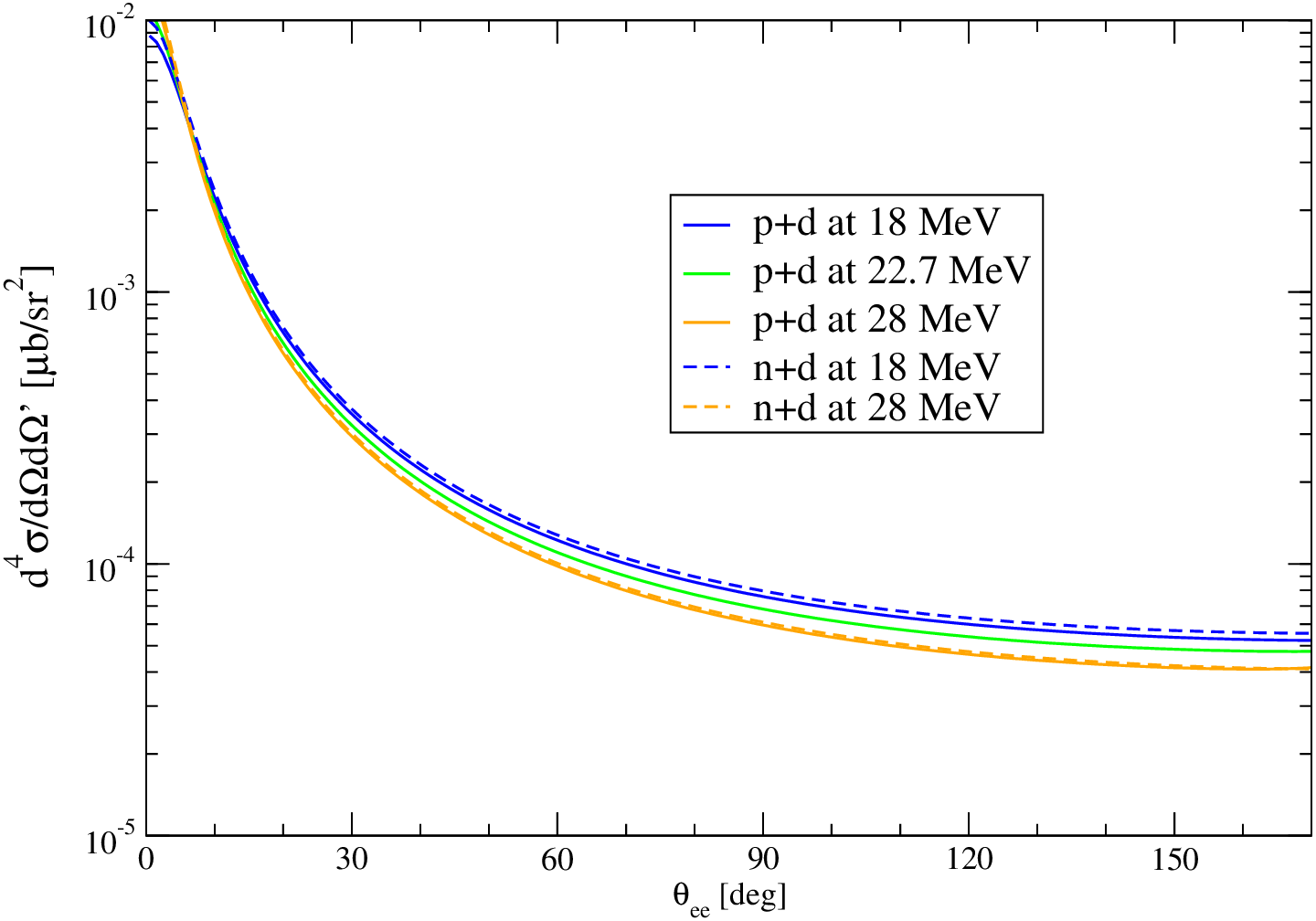}
\caption{(color online) The EM-only four-fold differential cross
  section for the $d(p,e^-e^+)\het$ and $d(n,e^-e^+)\tri$ processes
  calculated at different incident nucleon energies with the AV18UIX interaction
  (and accompanying EM currents); the kinematical configuration corresponds
to the lepton pair being emitted in the plane perpendicular
to the nucleon incident momentum, and $\theta_{ee}$ is the angle
between the electron and positron momenta. 
The cross sections
for the $d(p,e^-e^+)\het$ ($d(n,e^-e^+)\tri$) processes are shown as solid (dashed) lines.
The results for the NVIa3N interaction in practice overlap with those reported in the figure. }
\label{fig:sig4_EM}
\end{figure}

In Fig.~\ref{fig:sig4_EM} we show the $d(p,e^-e^+)\het$ and $d(n,e^-e^+)\tri$
four-fold differential cross sections calculated with the AV18UIX interaction,
corresponding to the kinematical configuration in which the lepton pair is
emitted in the plane perpendicular to the incident nucleon momentum
($\theta\,$=$\,\theta^\prime\,$=$\,90^\circ$) and as function of the relative
angle $\theta_{ee}$, that is, the angle between the electron and positron
momenta. As it can be seen, at each energy, the two cross sections essentially
overlap, a result we have anticipated.  The cross section has the typical form
for an IPC process dominated by the $E_1$ transition, i.e. it decreases monotocally
as $\theta_{ee}$ increases, becoming almost flat as $\theta_{ee}\rightarrow 180$ deg.
The small differences between the $p+d$ and $n+d$ results are due mainly to the different
value of $E_0$, namely the energy at disposal for the two leptons. In fact, in the $n+d$
process, there is the formation of a more bound trinucleon system ($\tri$).
Hence, the $d(n,e^-e^+)\tri$ cross section is slightly greater than that of  $d(p,e^-e^+)\het$.
The results obtained with the NVIa3N interaction are very similar to those shown
in Fig.~\ref{fig:sig4_EM}, in practice the curves obtained
with both interactions (and corresponding set of EM transition operators) essentially
overlap.

\begin{table*}[bth]
  \caption{\label{tab:totxs} Total cross sections (in $\mu$b) for the processes $d(p,e^-e^+)\het$,
    $d(p,\gamma)\het$, $d(n,e^-e^+)\tri$, and  $d(n,\gamma)\tri$ calculated at a number of incident
    nucleon energies $T_N$ (in MeV)  with the AV18UIX and NVIa3N Hamiltonians. }
  \begin{center}
    \begin{tabular}{c|cccc|cccc}
     \hline
     \hline
    & \multicolumn{4}{c}{AV18UIX} &  \multicolumn{4}{c}{NVIa3N}\\
     $T_N$ & $d(p,e^-e^+)\het$ & $d(p,\gamma)\het$  & $d(n,e^-e^+)\tri$ & $d(n,\gamma)\tri$ &
             $d(p,e^-e^+)\het$ & $d(p,\gamma)\het$  & $d(n,e^-e^+)\tri$ & $d(n,\gamma)\tri$ \\
     \hline
      $18.0$  & $0.0175$ & $9.06$ & $0.0192$  & $9.55$  & $0.0182$ & $9.39$ &  $0.0194$  & $9.59$ \\
      $22.7$  & $0.0165$ & $8.05$ & $0.0178$  & $8.35$  & $0.0170$ & $8.23$ &  $0.0177$  & $8.29$ \\
      $28.0$  & $0.0151$ & $7.02$ & $0.0159$  & $7.11$  & $0.0154$ & $7.18$ &  $0.0161$  & $7.19$ \\
      \hline
    \end{tabular}
    \end{center}
  \end{table*}

Total cross sections for the processes $d(p,e^-e^+)\het$,  $d(p,\gamma)\het$, $d(n,e^-e^+)\tri$, and  $d(n,\gamma)\tri$,
calculated at a number of incident  nucleon energies $T_N$ (in MeV)  with the AV18UIX and NVIa3N Hamiltonians (and accompanying currents)
are reported in Table~\ref{tab:totxs}.
We note that pair production cross sections are suppressed by a factor of approximately
$500$ relative to radiative capture cross sections. As expected, the difference between the $p+d$ and $n+d$
cross sections is small.

\subsection{Results including the X17 boson}
\label{sec:rmex}
Here we discuss the RMEs derived from the operators given in Eqs.~(\ref{eq:e94})--(\ref{eq:e99}).  The $q$ dependence of the RMEs
for either the $S$ or $V$ cases are similar to the behaviour already discussed for the EM current. However, now we can analyze the
behaviour of the RMEs originating from purely isoscalar and isovector operators. To be definite, we consider the
RMEs calculated for the $p+d$ process at $T_p=18$ MeV with the AV18UIX interaction.

\begin{figure}[bth]
\centering
 \includegraphics[scale=0.35,clip]{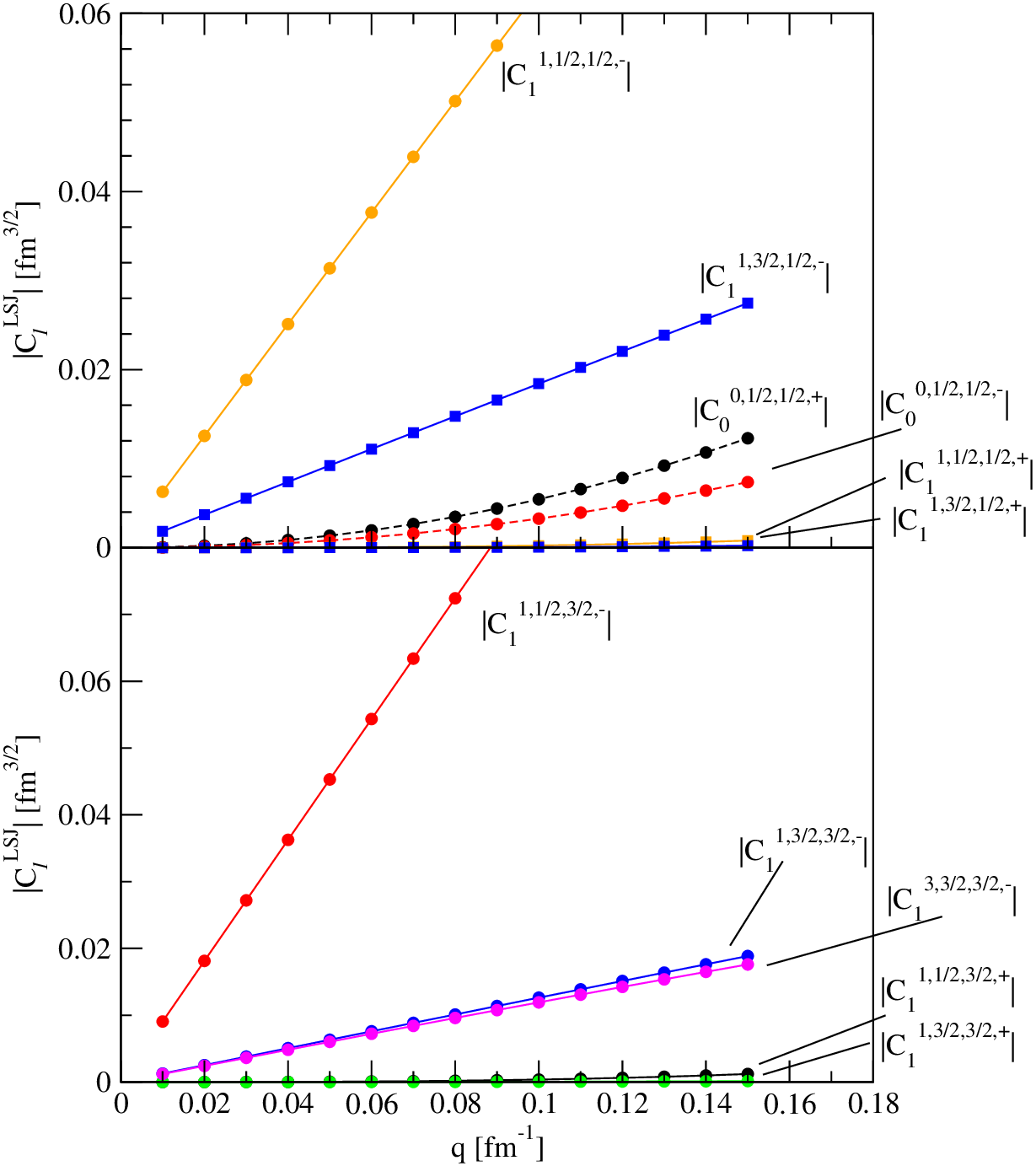}
\caption{(color online) The dependence on the three-momentum transfer $q$ of some RMEs for the
  charge operators given in Eq.~(\ref{eq:e94}); the calculations are at incident proton
  energy of 18 MeV and use the AV18UIX interaction.  
  The solid, dashed, an dotted lines show fits of the calculated values using linear, quadratic, and cubic parametrizations,
  respectively. The RMEs not shown in this plot are negligible.}
\label{fig:V_C}
\end{figure}

Let us consider first the $S$ case. The corresponding charge RMEs calculated using the operators given in Eq.~(\ref{eq:e94})
are shown in Fig.~\ref{fig:V_C}. As it can be seen, in case of the transitions
${1\over2}^-\longrightarrow{1\over2}^+$ and ${3\over2}^-\longrightarrow{1\over2}^+$, the
$C_1^{1,S,J,+}$ RMEs are dramatically suppressed. It is easy to show that they behave as $q^3$.
In fact, they are calculated from the matrix elements
\begin{equation}
  \langle \het | \sum_{j=1,3} e^{i\bmq\cdot\bmr^{(CM)}_j} | pd, {}^{2S+1}P_J\rangle \ ,
\end{equation}
where $\bmr^{(CM)}_j$ above is the distance of particle $j$ to the CM of the three particle system (the dependence
on the CM position $R_{CM}$ has been integrated out to obtain the momentum conservation). Expanding the plane wave,
the even powers of $(i\bmq\cdot\bmr^{(CM)}_j)$ vanish due to the parity constraint. The linear term in
$i\bmq\cdot\bmr^{(CM)}_j$ also vanishes since in the integral 
\begin{equation}
  \langle \het | \sum_{j=1,3} i\bmq\cdot\bmr^{(CM)}_j | pd, {}^{2S+1}P_J\rangle \ ,
\end{equation}
we have $\sum_j \bmr^{(CM)}_j=0$ by definition. The first nonvanishing term is therefore proportional to $q^3$.
Since in our case $q$ is small, the corresponding RMEs are suppressed. We note that the ``$\lambda=-$'' RMEs
come from the matrix element of the operator given in Eq.~(\ref{eq:e94}) 
\begin{equation}
  \langle \het | \sum_{j=1,3} e^{i\bmq\cdot\bmr^{(CM)}_j}\tau_{j,3} | pd, {}^{2S+1}P_J\rangle \ ,
\end{equation}
and consequently the first nonvanishing contribution is the linear one, proportional to $q$. 

\begin{figure}[bth]
\centering
 \includegraphics[scale=0.35,clip]{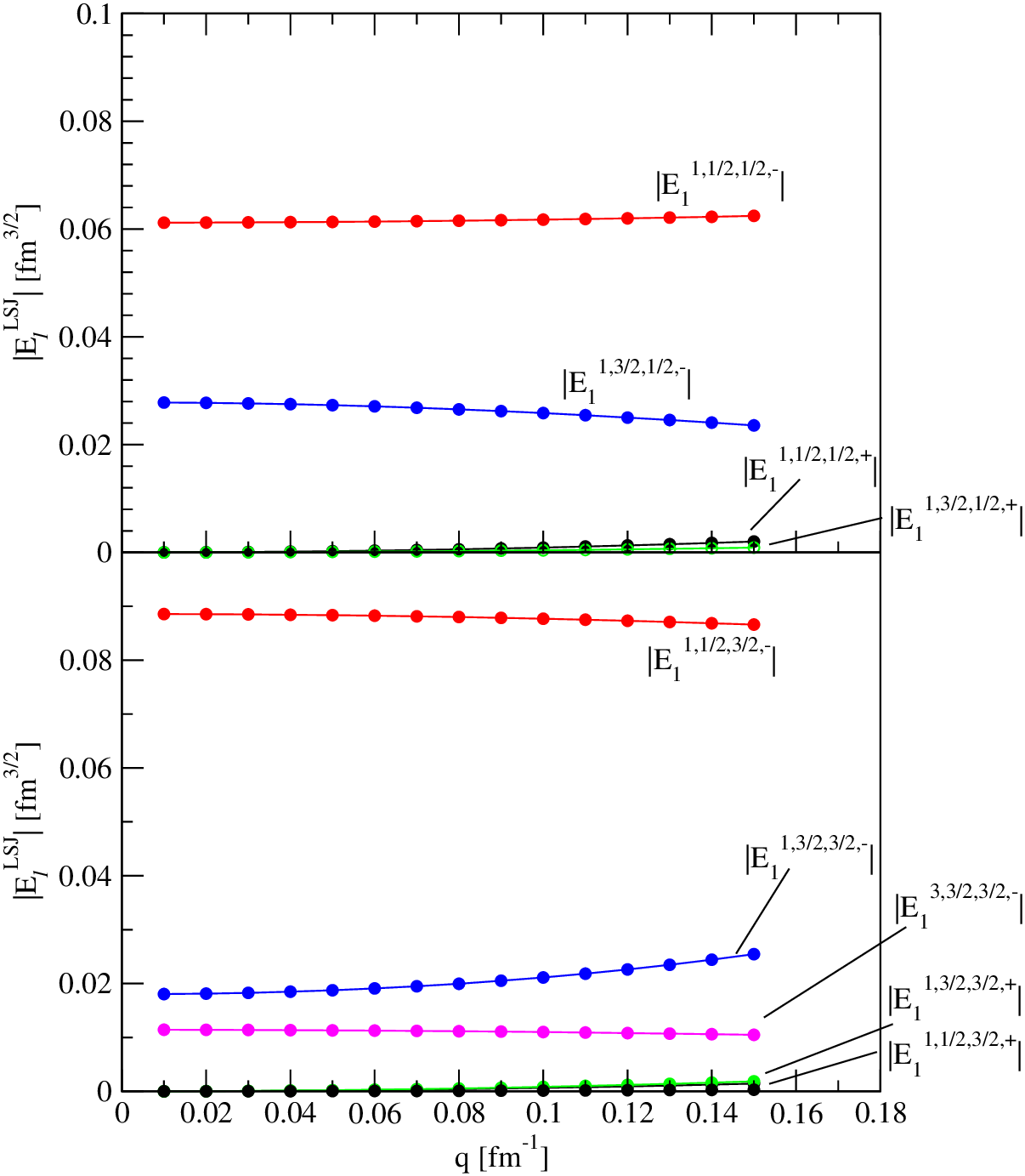}
\caption{(color online) The same as Fig.~\protect\ref{fig:V_C} but for the electric RMEs coming from Eq.~(\protect\ref{eq:e97}).}
\label{fig:V_E}
\end{figure}

In case of a vector X17, the charge RMEs derive from the same operators given in Eq.~(\ref{eq:e94}).
Regarding the transverse and longitudinal RMEs, we need to consider the operators given in
Eqs.~(\ref{eq:e97}) and~(\ref{eq:e98}). Some of the electric RMEs given by the operator of Eq.~(\ref{eq:e97}) are shown in Fig.~\ref{fig:V_E}.
Again, the ``$\lambda=+$'' components are suppressed, since this isoscalar operator can be written as
\begin{eqnarray}
 {\bm j}^{V+}(\bmq) &=& \frac{1}{2\, m_N} \sum_j \left[ e^{i\bmq\cdot\bmr^{(CM)}_j}\, , \,\bmp_j \right ]_+  \ ,\nonumber\\
  &\approx&  \frac{1}{2\, m_N} \sum_j \left[ \Bigl(1+{1\over2} \Bigl(i\bmq\cdot\bmr^{(CM)}_j\Bigr)^2\Bigr) , \,\bmp_j \right ]_+
     \ ,\nonumber\\
     &=&  \frac{1}{2\, m_N} \Bigl\{ 2\bmP \nonumber\\
     && \qquad + \sum_j \left[ {1\over2} \Bigl(i\bmq\cdot\bmr^{(CM)}_j\Bigr)^2 , \,\bmp_j \right ]_+
    \Bigr\}\ ,
\end{eqnarray}
where $\bmP=\sum_i \bmp_i$. The matrix element between the nuclear states  is then proportional to $q^2$, 
since $\bmP| pd, {}^{2S+1}P_J\rangle=0$ (the ket contains the CM wave function).
On the other hand, the same argument does not apply to the ``$\lambda=-$'' component, due
to the presence of the operator $\tau_{i,3}$ in the sum over the particles. In fact the ``$\lambda=-$'' RMEs are independent on $q$,
and therefore are much larger than the ``$\lambda=+$'' RMEs.

In the $V$ current,  we have added also the contribution
of the operator given in Eq.~(\ref{eq:e98}). In this case, the contribution of the ``$\lambda=+$'' and ``$\lambda=-$'' components are of the
same order of magnitude. However, this contribution is much less than the contribution of $j^V$, and therefore
this term does not change the situation. This is confirmed by the calculations, reported in Fig.~\ref{fig:V_E}.
The longitudinal RMEs behave similarly, while the magnetic RMEs are always very small. 

\begin{figure}[bth]
\centering
 \includegraphics[scale=0.35,clip]{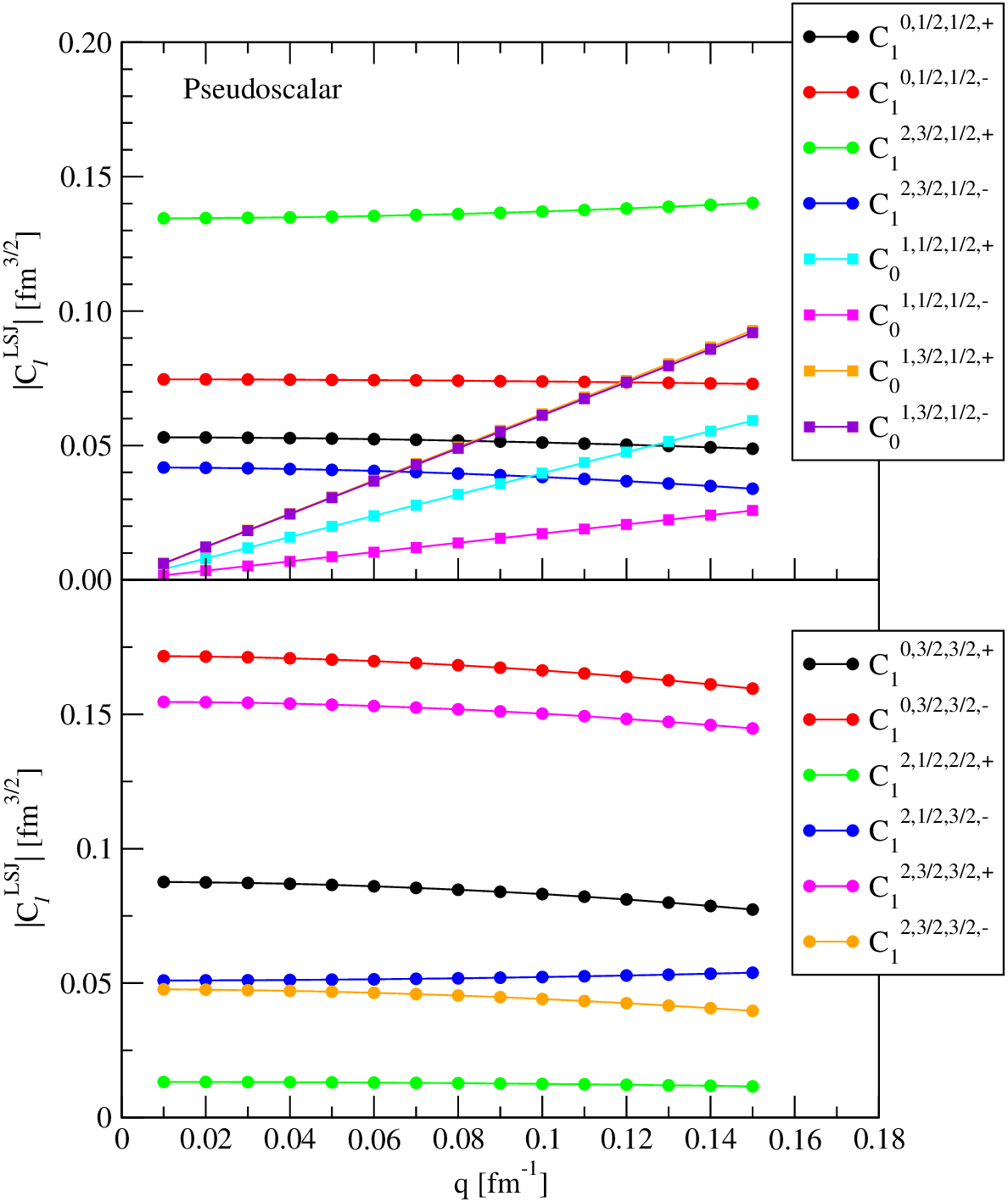}
\caption{(color online) Same as Fig.~\protect\ref{fig:V_C} but for the charge RMEs derived from the
pseudoscalar operator given in Eq.~(\ref{eq:e95}).}
\label{fig:X_P}
\end{figure}

\begin{figure}[bth]
\centering
 \includegraphics[scale=0.35,clip]{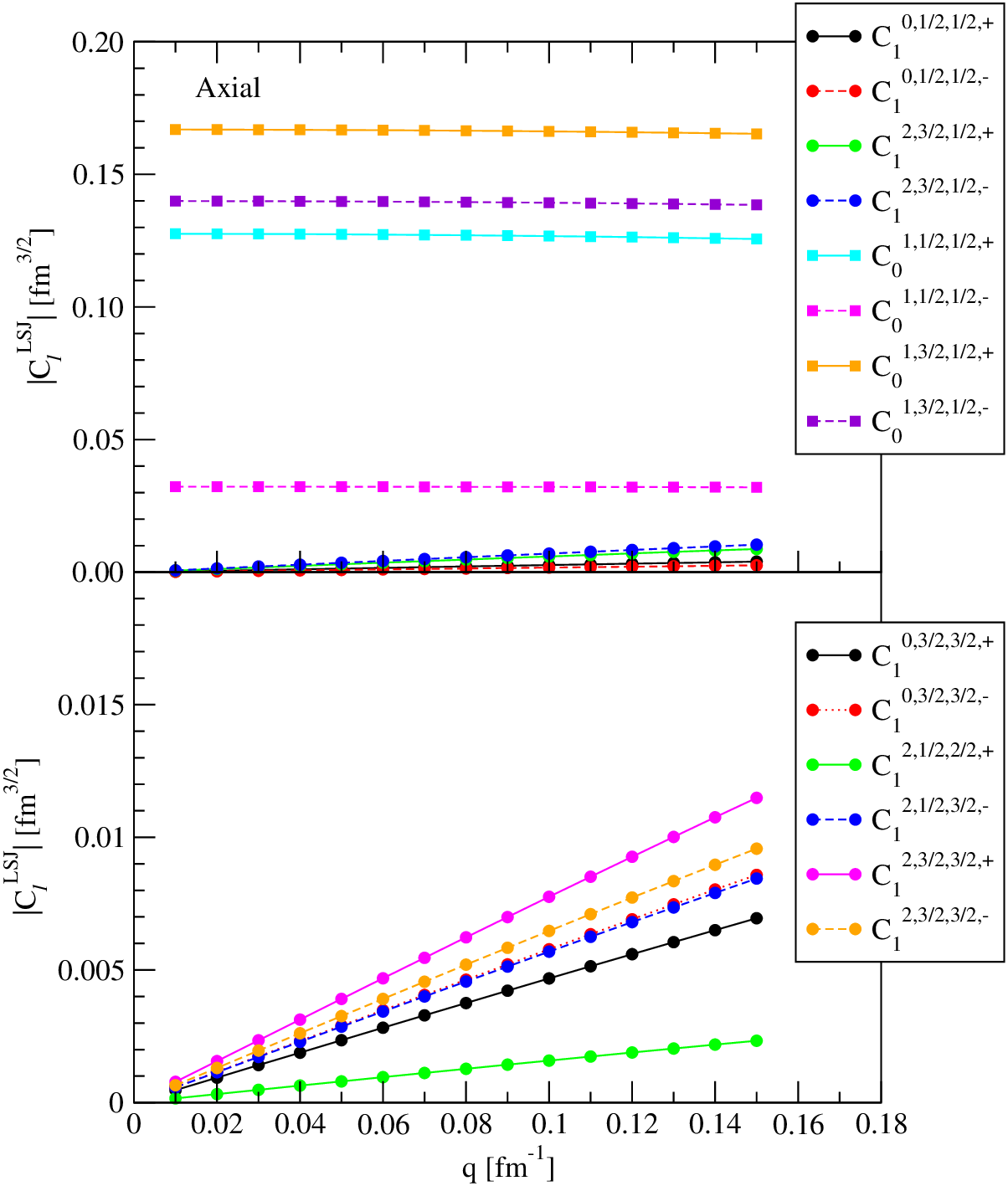}
\caption{(color online) 
  Same as in Fig.~\ref{fig:V_C} bur for the charge RMEs derived from the
 axial operator given in Eq.~(\ref{eq:e97}).}
\label{fig:X_AC}
\end{figure}

\begin{figure}[bth]
\centering
 \includegraphics[scale=0.35,clip]{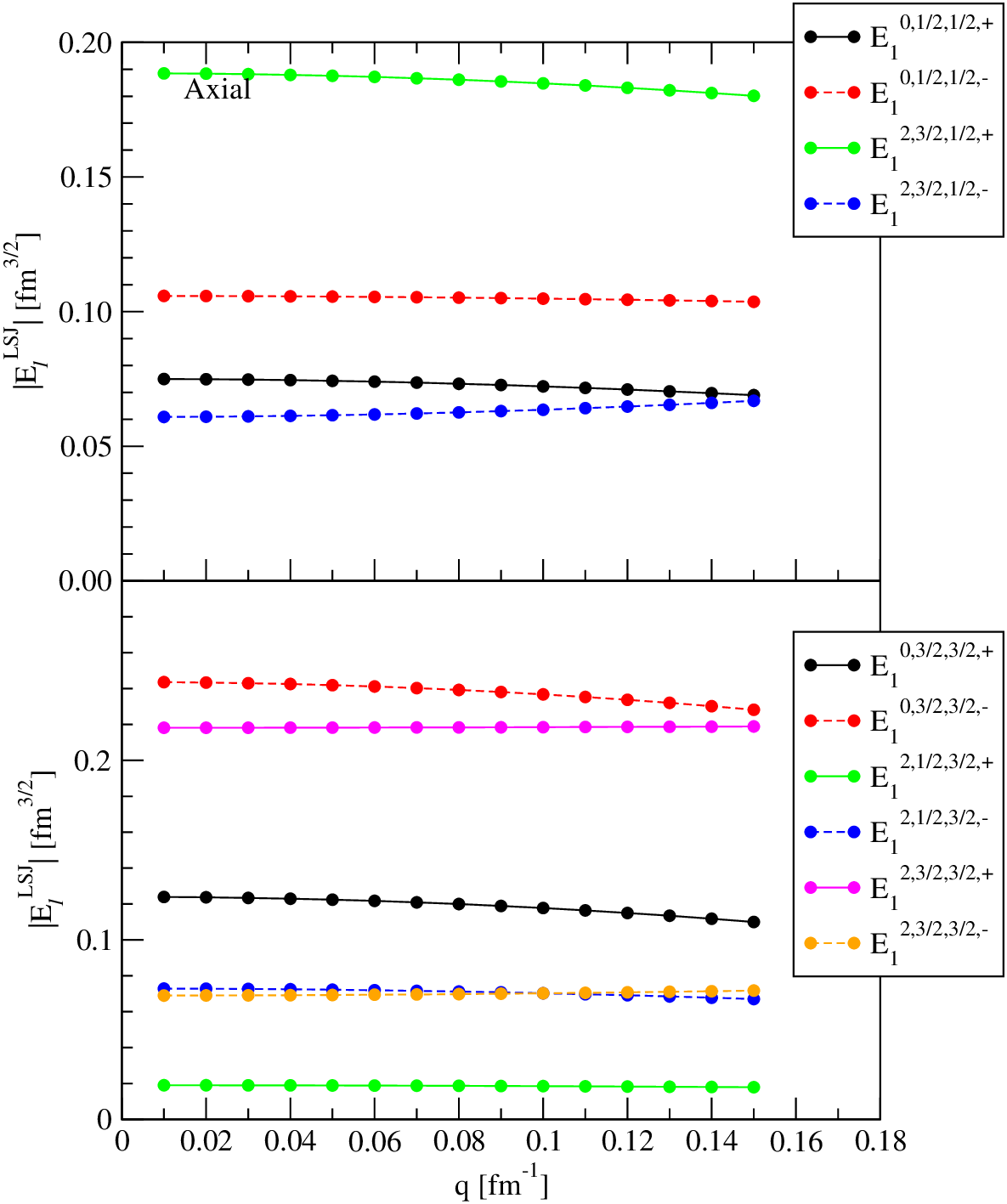}
\caption{(color online) 
  Same as in Fig.~\ref{fig:V_C} bur for the electric RMEs derived from the
  axial operator given in Eq.~(\ref{eq:e99}).}
\label{fig:X_AE}
\end{figure}

The $q$ dependence of the RMEs coming from the $P$ and $A$ operators reported in Eq.~(\ref{eq:e95}),
~(\ref{eq:e96}), and~(\ref{eq:e99}) are shown in Figs.~\ref{fig:X_P},~\ref{fig:X_AC}, and ~\ref{fig:X_AE}.
Note that the RMEs associated with the pseudoscalar and the time component of the
axial operators behave differently. In these cases, there is no suppression of the ``$\lambda=+$'' component.

Regarding the energy dependence, in all cases the RMEs smoothly decrease as the energy is increased,
similar to what found for the EM RMEs.

The absolute values of a selected set of RMEs (generally the largest ones),
calculated for the $n+d$ and $p+d$ processes, are compared in Table~\ref{tab:pdndX}.
For the $S$ and $V$ cases, we report the dominating $C_1$ and $E_1$ RMEs, coming from the
transitions ${1\over2}^-\longrightarrow{1\over2}^+$ and ${3\over2}^-\longrightarrow{1\over2}^+$.
As it can be seen, the difference between the $p+d$ and $n+d$ absolute values of these RMEs are rather tiny.
We note again that the RMEs originating from the ``$\lambda=+$'' operators are much smaller than those
calculated from the ``$\lambda=-$'' operators. On the other hand, for the $P$ and $A$ cases, there is no
particular difference between the ``$\lambda=+$'' and ``$\lambda=-$'' RMEs. Regarding the effective values of the $p+d$ and $n+d$ RMEs
(not the absolute values), generally the ``$\lambda=+$'' RMEs have the same sign, while the ``$\lambda=-$'' have different sign.

\begin{table}[bth]
  \caption{\label{tab:pdndX}
A selected set of RMEs in absolute value (in fm$^{3/2}$) corresponding to the $p+d$ and $n+d$ reactions,
obtained with the AV18UIX interaction and the operators given in Eqs.~(\ref{eq:e94})--(\ref{eq:e99}).
The incident nucleon energy is 18 MeV and the three-momentum transfer $q$ is 0.1 fm$^{-1}$.}  
  \begin{center}
    \begin{tabular}{lccccc}
            \hline
            \hline
      \multicolumn{6}{c}{S case}\\
      \hline 
      RMEs$\times 10^3$ & wave & \multicolumn{2}{c}{$p+d$} &  \multicolumn{2}{c}{$n+d$} \\
      && $\lambda=+$ & $\lambda=-$ & $\lambda=+$ & $\lambda=-$ \\
      \hline
      $|C_1^{1,{1\over2},{1\over2},\lambda}|$ & ${}^2P_{1\over2}$ &  $0.22$ & $62.2$ & $0.19$ & $61.0$ \\
      $|C_1^{1,{1\over2},{3\over2},\lambda}|$ & ${}^2P_{3\over2}$ &  $0.33$ & $90.9$ & $0.29$ & $88.0$ \\
      \hline
      \multicolumn{6}{c}{V case}\\
      \hline 
      RMEs$\times 10^3$ & wave & \multicolumn{2}{c}{$p+d$} &  \multicolumn{2}{c}{$n+d$} \\
      && $\lambda=+$ & $\lambda=-$ & $\lambda=+$ & $\lambda=-$ \\
      $|E_1^{1,{1\over2},{1\over2},\lambda}|$ & ${}^2P_{1\over2}$ &  $0.90$ & $61.7$ & $0.87$ & $61.8$ \\
      $|E_1^{1,{3\over2},{1\over2},\lambda}|$ & ${}^4P_{1\over2}$ &  $0.40$ & $25.8$ & $0.40$ & $26.9$ \\
      $|E_1^{1,{1\over2},{3\over2},\lambda}|$ & ${}^2P_{3\over2}$ &  $0.64$ & $87.7$ &$0.62$ & $87.8$ \\
      $|E_1^{1,{3\over2},{3\over2},\lambda}|$ & ${}^4P_{3\over2}$ &  $0.82$ & $21.1$ &$0.82$ & $21.8$ \\
      \hline
      \multicolumn{6}{c}{P case}\\
      \hline 
      RMEs$\times 10^3$ & wave & \multicolumn{2}{c}{$p+d$} &  \multicolumn{2}{c}{$n+d$} \\
      && $\lambda=+$ & $\lambda=-$ & $\lambda=+$ & $\lambda=-$ \\
      $|C_1^{0,{1\over2},{1\over2},\lambda}|$ & ${}^2S_{1\over2}$ &  $51.1$ & $73.8$ &$50.4$ & $69.6$ \\
      $|C_1^{0,{3\over2},{3\over2},\lambda}|$ & ${}^4S_{3\over2}$ &  $83.1$ & $166.3$ &$83.2$ & $156.9$ \\
      \hline
      \multicolumn{6}{c}{A case}\\
      \hline 
      RMEs$\times 10^3$ & wave & \multicolumn{2}{c}{$p+d$} &  \multicolumn{2}{c}{$n+d$} \\
      && $\lambda=+$ & $\lambda=-$ & $\lambda=+$ & $\lambda=-$ \\
      $|C_0^{1,{1\over2},{1\over2},\lambda}|$ & ${}^2P_{1\over2}$ &  $127.7$ & $32.1$ &$127.1$ & $33.4$ \\
      $|C_0^{1,{3\over2},{1\over2},\lambda}|$ & ${}^4P_{1\over2}$ &  $166.2$ & $139.3$ &$168.9$ & $140.1$ \\      
      $|E_1^{0,{1\over2},{1\over2},\lambda}|$ & ${}^2S_{1\over2}$ &  $72.2$ & $104.8$ &$71.2$ & $98.7$ \\
      $|E_1^{0,{3\over2},{3\over2},\lambda}|$ & ${}^4S_{3\over2}$ &  $117.7$ & $236.8$ &$117.9$ & $223.3$ \\      
      \hline
      \hline
    \end{tabular}
    \end{center}
  \end{table}

Let us now present the results of the four-fold cross section for the processes $d(p,e^+ e^-)\het$ and $d(n,e^+ e^-)\tri$.
The cross sections have been calculated assuming the values for
$\varepsilon_0^c$ and $\varepsilon_z^c$  given in Table~\ref{tab:eta_values}. These values were obtained in
Ref.~\cite{Viviani:2021stx} by reproducing the 2019 ATOMKI data for the process $\tri(p,e^+ e^-)\heq$,
using the same NVIa3N interaction employed here. However,
as discussed in that paper, such extraction is rather uncertain, due to the quality of those experimental data
(lack of subtraction of the leptonic pairs produced by real gammas hitting the apparatus, etc.). So these values
have to be considered only as indicative. Anyway, we will use them in this paper to have
an idea of the eventual X17 effects in the $A=3$ system.

\begin{table}[bth]
\caption{\label{tab:eta_values}
Values of the coupling constants $\varepsilon_0^c$ and $\varepsilon_z^c$, $c=S,P,V,A$,  obtained in Ref.~\protect\cite{Viviani:2021stx} from the
fit of the $\tri(p,e^+e^-)\heq$ 2019 ATOMKI angular distribution~\protect\cite{Krasznahorkay:2019lyl} at $T_p=0.90$ MeV. In all cases,
one of the coupling constant has been fixed to a given value (evidentiated in bold), while the other
has been chosen in order to reproduce the ATOMKI data. In the V case, the adopted choice corresponds
to a proto-phobic X17. See Ref.~\protect\cite{Viviani:2021stx} for more details. }
\begin{center}
\begin{tabular}{l|cc}
\hline 
\hline
 & \multicolumn{2}{c}{NVIa3N}\\
\hline 
Case $c$ &  $\varepsilon_0^c$  & $\varepsilon_z^c$ \\
\hline
$S$  &  $0.75\times 10^0$ & ${\bm 0}$ \\
$P$  &  $2.72\times 10^1$ & ${\bm 0}$  \\
$V$  &   $2.66\times 10^{-3}$ & $-{\bm 3}\,{\bm  \varepsilon}_0^V$ \\
$A$  &  $2.89\times10^{-3}$ & ${\bm 0}$ \\
\hline 
\hline 
\end{tabular}
\end{center}
\end{table}

\begin{figure}[bth]
\centering
 \includegraphics[scale=0.35,clip]{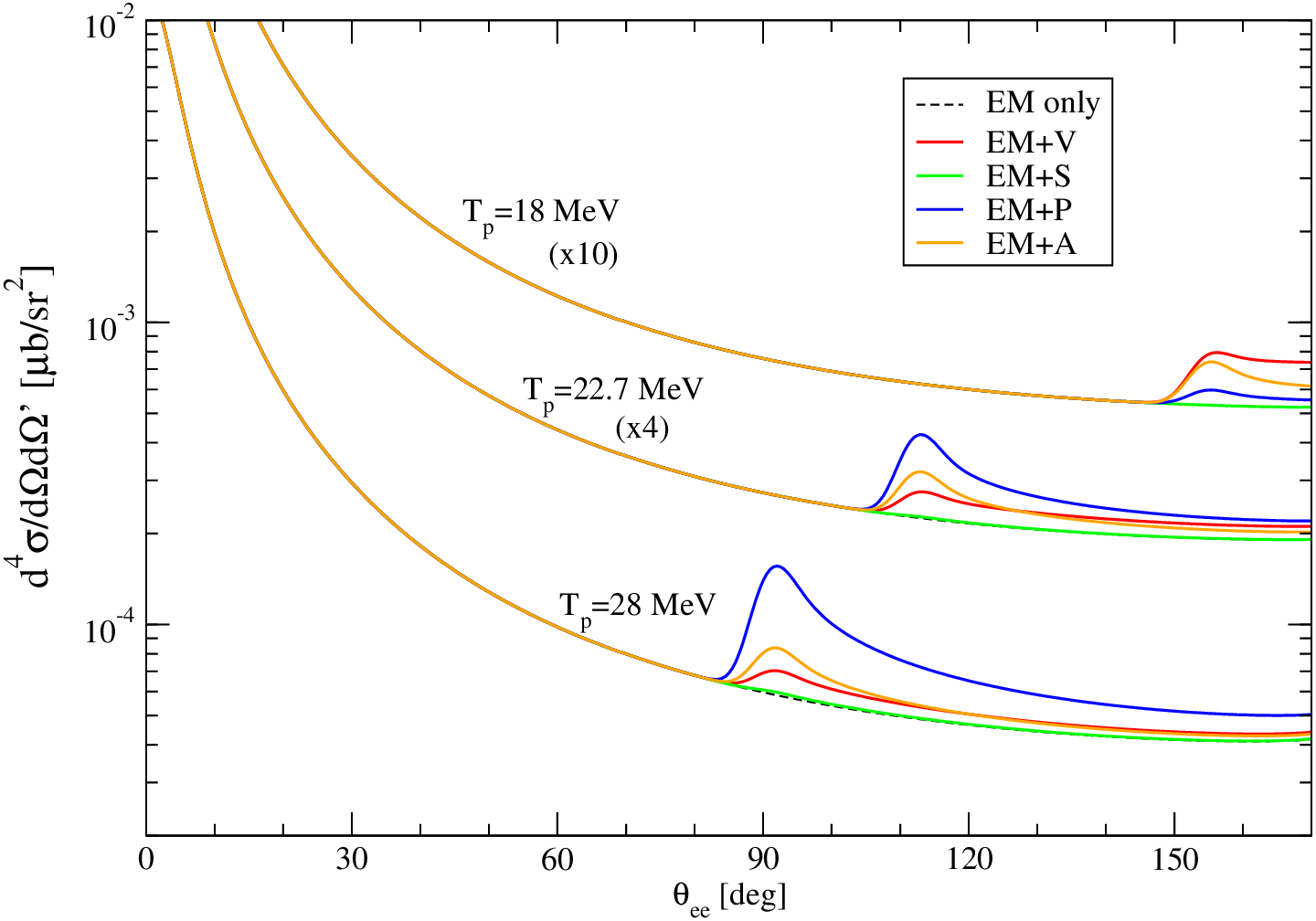}
\caption{(color online) 
  The $d(p,e^+ e^-)\het$  four-fold cross sections calculated with the NVIa3N interaction as function of the angle $\theta_{ee}$ at various energies. Here $\theta=\theta'=90$ deg. The
  EM-only cross section (dashed lines) is compared with those obtained by including the effect of the exchange of an X17,
  for the S, P, V, and A cases. The used quark-X17 coupling constants are given in Table~\protect\ref{tab:eta_values}. The EM+S curve
  is always almost coincident to the EM curve. Note that the cross sections calculated at $T_p=18$ MeV
  ($22.7$ MeV) have been rescaled by a factor 10 (4) for the sake of clarity. As discussed at the end of Sec.~\protect\ref{sec:theo},
  the cross section has been folded with a normalized Gaussian of width $\Delta=2.5$ deg,
  in order to approximately take into account of experimental finite angular resolution of the detectors.
}
\label{fig:sig4_pd_X}
\end{figure}

Once clarified this point, the $d(p,e^+ e^-)\het$  four-fold cross sections calculated with NVIa3N at various energies
are reported in Fig.~\ref{fig:sig4_pd_X}, for the emission of the lepton in the laboratory plane
perpendicular to the incident beam momentum ($\theta=\theta'=90$ deg), and as function of the angle
$\theta_{ee}$. Several comments are in order. (i) The X17 peak moves to lower values of $\theta_{ee}$ as the energy increases,
as already discussed in Sec.~\ref{sec:kine}. Note the correspondence between the minimum angles where the
X17 signal appears shown in Fig.~\ref{fig:ytheta} and the position of the peaks in
Fig.~\ref{fig:sig4_pd_X}. (ii) For the scalar case, the X17 contribution is very tiny, and the
cross section almost coincident with the EM cross section only. This is due to the fact that for the $\tri(p,e^+ e^-)\het$ process
at $T_p=0.90$ MeV we are rather close to a $0^+$ $\heq$ resonance, and therefore the X17 RMEs in this case are large,
and the adopted value of $\varepsilon^S_0$ sufficiently small to reproduce the observed peak. In the present case,
however, the RMEs induced by the $S$ operators are small, and therefore the peak is almost unobservable.
(iii) For the pseudoscalar case, the peak becomes more and more evident as the beam energy increases. In the case under
study ($\varepsilon^P_0\ne0$ and $\varepsilon^P_z=0$), the transition operator is proportional to $q$, whose value $q_{\text peak}$
at the peak increases as $E_0$ increases. In fact, for the $d(p,e^+e^-)\het$ we have $q_{\text peak}=4.1$, $11.7$, and $17.2$ MeV/c
for $T_p=18$, $22.7$, and $28$ MeV, respectively. This explains the rise of the peak in this case. 
(iv) The height of the peaks with respect to the EM-only cross section due to the vector X17 (at $T_p=22.7$ and $28$ MeV) is approximately constant with energy. This is related to the fact that the
operators inducing the X17 in this case are practically the same as those appearing in the EM current.
(v) The peaks at $T_p=18$ MeV are a bit at variance with respect to the other two energies, but this may be due
to the smallness of the phase space in this case.

So although the three nucleon spectrum has not any structure, it can be exploited to distinguish between the various cases.
(i) No observation of the signal with respect to the signal observed in $\tri(p,e^+ e^-)\het$ would be explained by
a scalar X17; (ii) a constant height of the peak when the beam energy is increased would indicate
the presence of a vector X17; (iii) an increase of the peak height with the beam energy would point out
to a pseudoscalar X17; (iv) the vector and axial cases could in principle be distinguished by looking
at the angular dependence of the cross section: as $\theta_{ee}\rightarrow180$ deg, the cross section
decays faster in the $A$ case than in the $V$ case. 

\begin{figure}[bth]
\centering
 \includegraphics[scale=0.35,clip]{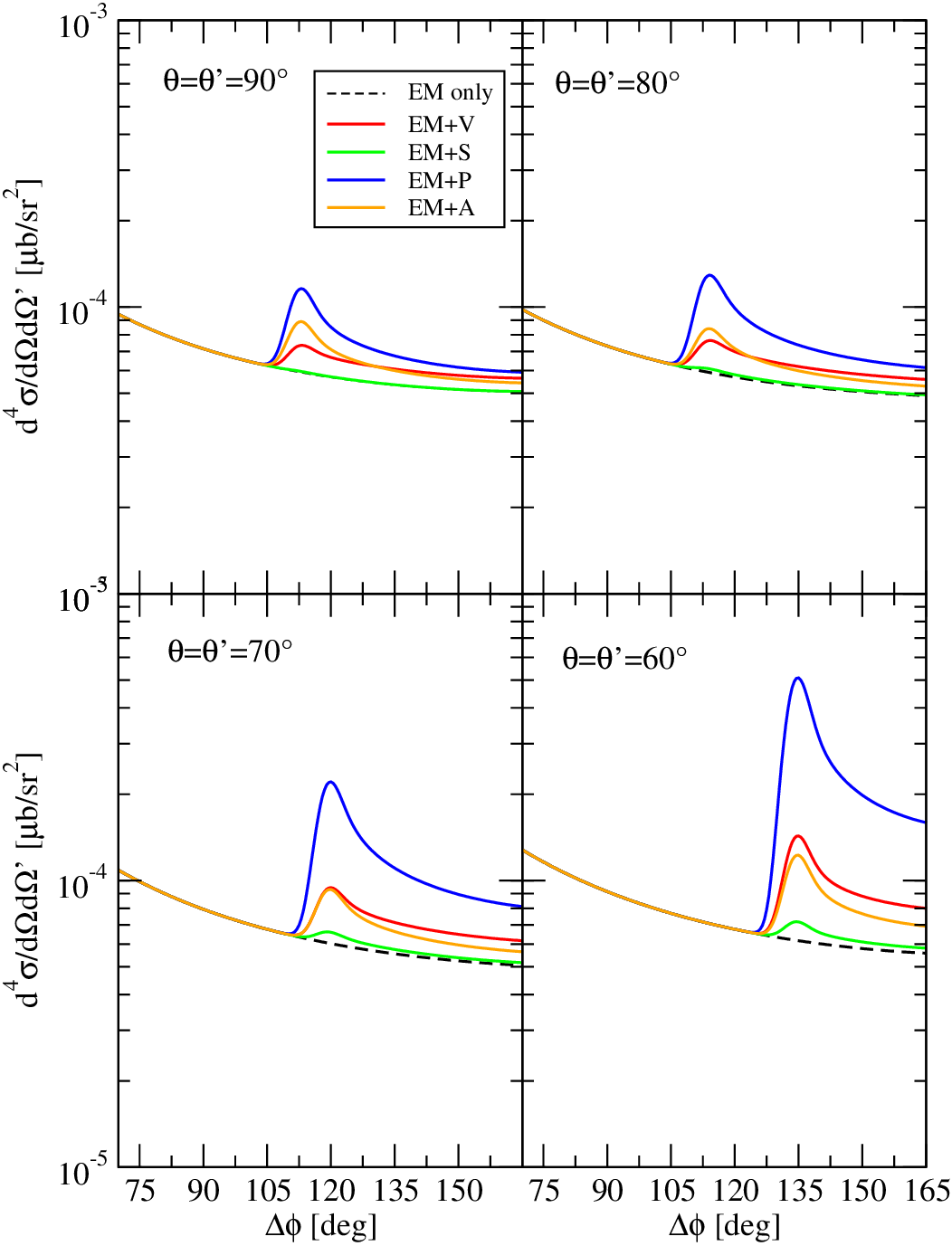}
\caption{(color online) 
  The $d(p,e^+ e^-)\het$ four-fold cross sections calculated with the NVIa3N interaction at $T_p=22.7$ MeV
  and for different values of $\theta$ and $\theta'$ (but retaining $\theta=\theta'$), the angles of the momenta of the two leptons with respect to the $z$ axis
  (the $z$ axis is chosen to be along the direction of the incident beam). The cross sections are reported
  as function of the angle $\Delta\phi=\phi-\phi'$ (the azimuthal angles of the two leptons).
  For the notation, see Fig.~\protect\ref{fig:sig4_pd_X}.
}
\label{fig:sig4_pd_X_theta}
\end{figure}

Clearly, a greater amount of information can be obtained by detecting the leptons out of the perpendicular plane. 
As an example, we report in Fig.~\ref{fig:sig4_pd_X_theta} the $d(p,e^+ e^-)\het$ four-fold cross section calculated with NVIa3N 
at $T_p=22.7$ MeV, for four different values  of the polar angles $\theta=\theta'$ (chosen to be equal). 
The cross section is calculated as function of $\Delta\phi=\phi-\phi'$ (the azimuthal angles of the two leptons).
For the $P$ case, we observe a noticeable increase of the height of the peak as $\theta=\theta'$ decrease. This happens
since at $\theta=\theta'=90$ deg the X17 contribution for this case is kinematically suppressed,
Then, out of the perpendicular plane, where the kinematical suppression is less and less effective,
the X17 contribution for the $P$ case starts to become larger and larger. 
For the $V$ and $A$ cases, the X17 contribution is approximately the same as $\theta=\theta'$ decrease. However, it is possible
to note that the peak for the $A$ case becomes smaller than the peak for the $V$ case as the lepton polar angles become
smaller and smaller. Finally, for $\theta=\theta'=60$ deg, the X17 contribution for the $S$ case starts to become 
visible, although it remains rather small. Therefore, from this example, we see that the X17 peak varies differently 
for the $S$, $P$, $V$, and $A$ cases as the polar angles are varied, giving better 
handles to learn about the X17 spin and parity. 

\subsection{The cross section for the $p+d$ and $n+d$ processes}
\label{sec:pdnd}

In this subsection, we investigate the difference of the X17 signal in the two processes
$d(p,e^+ e^-)\het$ and $d(n,e^+ e^-)\tri$.  Since in the second process
the number of neutrons is twice than in the first process, from the (eventual)
measurements of both cross sections, one could extract information about
the isospin dependence of the X17-nucleon interaction, namely the relative
magnitude of the coupling constants $\varepsilon_0^c$ and $\varepsilon_z^c$.
In particular, one could test the proto-phobic X17 hypothesis which, as discussed in the Sec.~\ref{sec:intro},
was formulated for a vector X17 in order to avoid the constraint of the NA48 limit. Hereafter we will not
consider the scalar case, as we have seen that the eventual X17 signal is expected to be very tiny
in these two reactions. 

\begin{figure}[bth]
\centering
 \includegraphics[scale=0.35,clip]{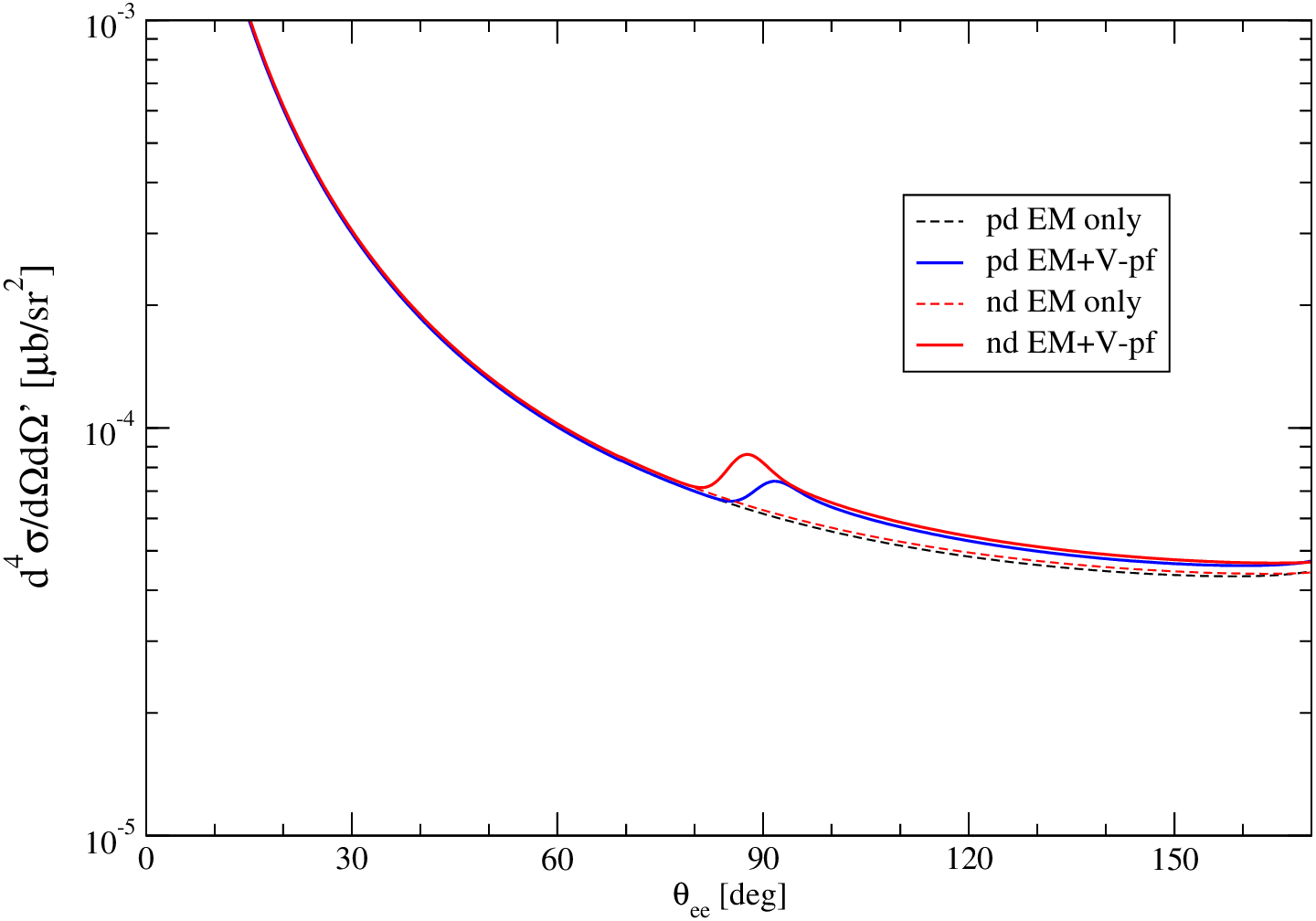}
\caption{(color online) 
  The $d(p,e^+ e^-)\het$ and $d(n,e^+ e^-)\tri$ four-fold cross sections calculated with the NVIa3N 
  interaction in function of the angle $\theta_{ee}$ at
  a beam energy of $28$ MeV ans for $\theta=\theta'=90$ deg.
  The cross sections for the two processes in case of a proto-phobic vector X17 (curves labeled EM+V-pf) are shown,
  and compared with those obtained for the EM-only case.
  As in Fig.~\protect\ref{fig:sig4_pd_X},
  the cross section has been folded with a normalized Gaussian of width $\Delta=2.5$ deg.}
\label{fig:sig4_pd_nd_V}
\end{figure}

Let us first consider the case of a vector X17. The cross sections for the
$d(p,e^+ e^-)\het$ and $d(n,e^+ e^-)\tri$ processes calculated at $T_N=28$ MeV
for a proto-phobic X17, i.e. with the coupling constants satisfying the condition of Eq.~(\ref{eq:protophobic}),
are shown in Fig.~\ref{fig:sig4_pd_nd_V}. As it can be seen, in this case, the heights of the two peaks are
comparable (for the $n+d$ process, the phase space is slightly larger and
the peak starts at lower values of $\theta_{ee}$, both effects due to the  
larger $E_0$). 

The vector current with the choice of Eq.~(\ref{eq:protophobic}) reads at LO
\begin{equation}
  \bmj_V=e\, \sum_i \varepsilon_0^V {1-\tau_{i,3}\over 2} e^{i\bmq\cdot\bmr_i^{(CM)}}+\ldots\  .\label{eq:jv}
\end{equation}
Now, for this case and in this range of energies, the largest RMEs are the $E_1$ RMEs, coming from the transitions
${1\over2}^-\longrightarrow{1\over2}^+$ and ${3\over2}^-\longrightarrow{1\over2}^+$. As we have already discussed
in the previous subsection, however, the isoscalar part of the operator in Eq.~(\ref{eq:jv}) gives vanishing contributions
with respect to the isovector part, and consequently $ \bmj_V|n+d\rangle\approx - \bmj_V|p+d\rangle$.
So the effect of a greater number of neutrons in the $d(n,e^+ e^-)\tri$ process is
not effective in this case. 

In the vector current we have also considered the part proportional to $\bmsi\times\bmq$. Both isoscalar and isovector
contributions are of the same order for this operator. However, the contribution to the RMEs coming from this term is rather small,
and therefore does not help. Further contributions at NLO come from one-pion-exchange diagrams, which however give isovector operators.
Clearly, matrix elements of these operators are of the same order of magnitude in the two processes.
Therefore, also these contributions will likely not give substantial differences between the $p+d$ and $n+d$ processes. . 

\begin{figure}[bth]
\centering
 \includegraphics[scale=0.35,clip]{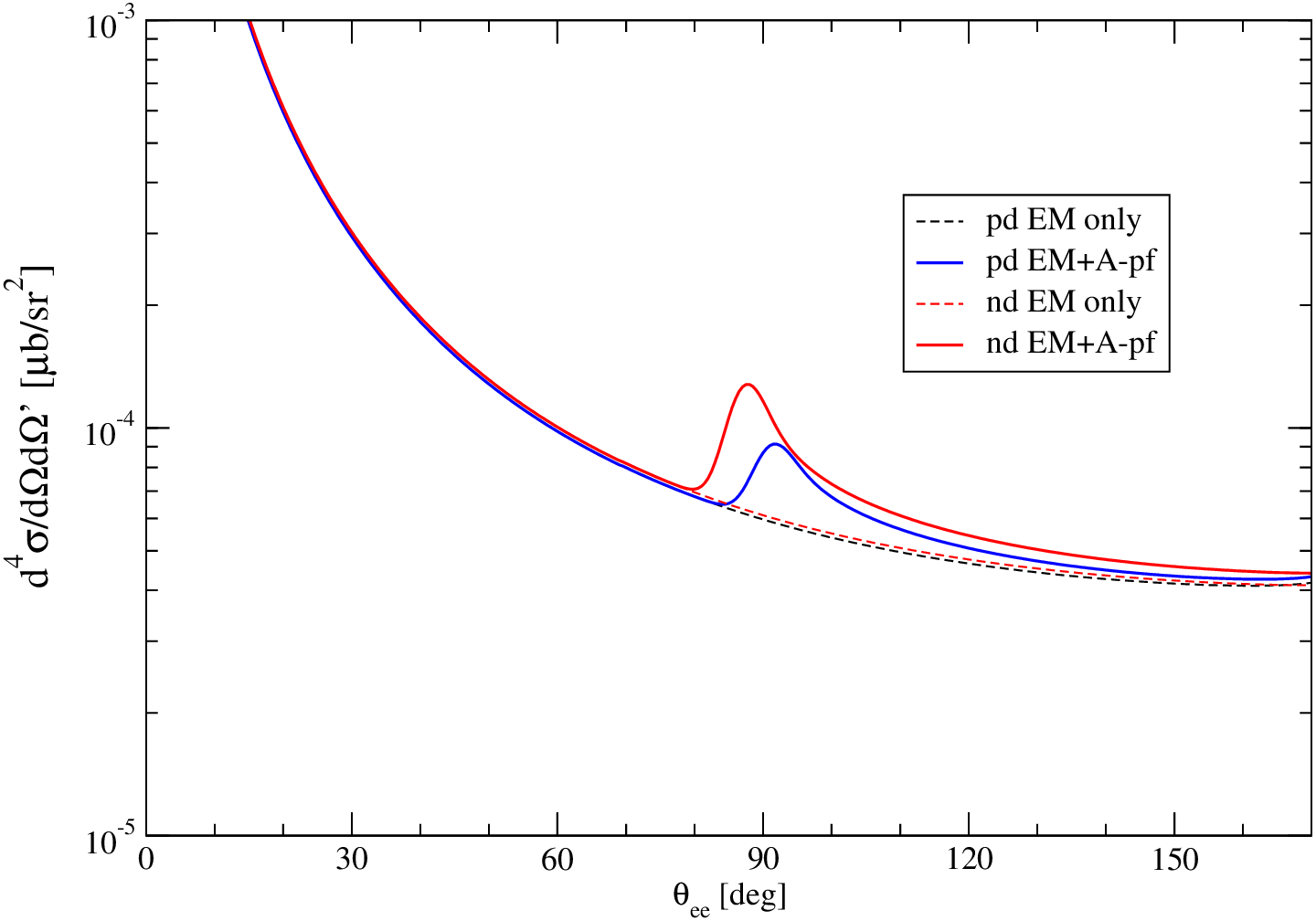}
\caption{(color online) 
  The same as in Fig.\protect\ref{fig:sig4_pd_nd_V} but for a proto-phobic axial X17 (curves labeled EM+A-pf).}
\label{fig:sig4_pd_nd_A}
\end{figure}

Let us consider the axial case. The dominant contribution comes from the spatial part, which in
our LO approximation reads
\begin{equation}
  \bmj_A=e\, \sum_i \Bigl[(3\,F-D)\varepsilon^A_0 +(F+D) \varepsilon^A_z\tau_{i,3}\Bigr] \bmsi_i e^{i\bmq\cdot\bmr_i^{(CM)}}\ .,
  \label{eq:ja}
\end{equation}
where we have used the expressions for $\eta_0^A$ and $\eta_z^A$ given in Eq.~(\ref{eq:cvp}). 
Let us choose a proto-phobic X17, namely so that $(3\,F-D)\varepsilon^A_0 +(F+D) \varepsilon^A_z=0$. The cross sections for the
$d(p,e^+ e^-)\het$ and $d(n,e^+ e^-)\tri$ processes calculated at $T_N=28$ MeV in this case 
are shown in Fig.~\ref{fig:sig4_pd_nd_A}. As it can be seen, the peak for the $n+d$ process is almost twice higher than for the
$p+d$ case. As we have seen in Sec.~\ref{sec:rmex}, in this case both the isoscalar and
isovector RMEs are of the same order of magnitude. Therefore, the $n+d$ cross section turns out to be larger than for the $p+d$ process. 

\begin{figure}[bth]
\centering
 \includegraphics[scale=0.35,clip]{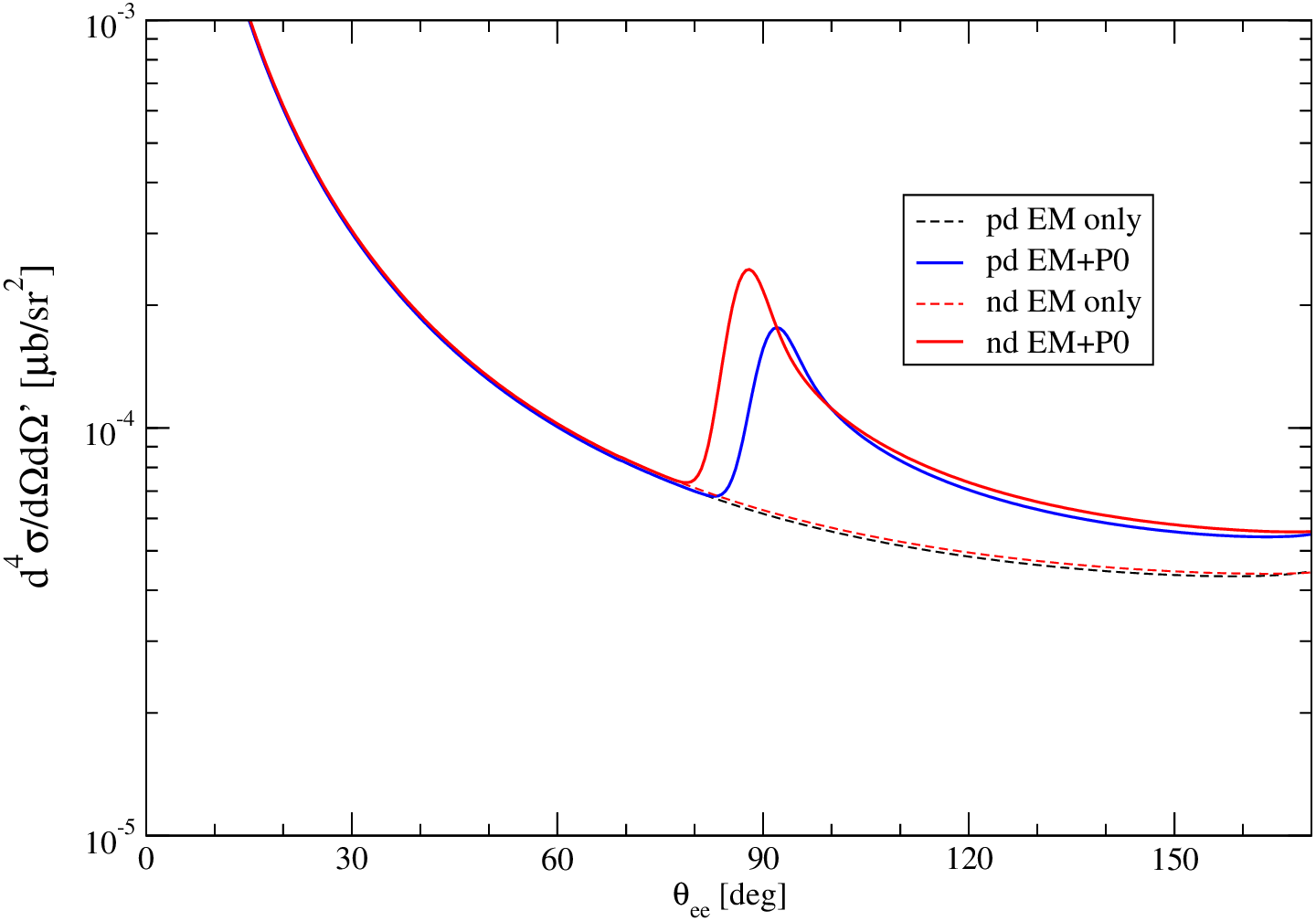}
\caption{(color online) 
  The same as in Fig.\protect\ref{fig:sig4_pd_nd_V} but for a isoscalar pseudoscalar X17 (curves labeled EM+P0).}
\label{fig:sig4_pd_nd_P}
\end{figure}

Finally, let us comment about the pseudoscalar case. As discussed in Sec.~\ref{sec:rmex}, the isoscalar and isovector operator have different
origin: the isoscalar comes from the direct coupling of X17 to the nucleon, coming by terms in the N2LO Lagrangian and proportional
to the LECs $d_{18}+2 d_{19}$. The isovector operator comes from the direct coupling of the X17 to the neutral pion. Therefore,
it would be rather strange that the different coupling would be fine-tuned to create a proto-phobic interaction.
According to the model of Refs.~\cite{Alves:2017avw,Alves:2020xhf}, the
direct pion-X17 is suppressed. Therefore, we have (at the chiral order considered in this paper) only an isoscalar interaction. Considering this latter case,
the cross sections for the $d(p,e^+ e^-)\het$ and $d(n,e^+ e^-)\tri$ processes calculated at $T_N=28$ MeV with NVIa3N
are shown in Fig.~\ref{fig:sig4_pd_nd_P}. As it can be seen, the two cross sections have similar magnitude, as expected for
an interaction which does not distinguish between protons and neutrons. The slightly larger $n+d$ cross section is related
to the phase space (larger value of $E_0$). 

\section{Conclusions}
\label{sec:conc}

In this paper, we have studied the $e^+$-$e^-$ pair production in the
$d(p,e^+ e^-)\het$ and $d(n,e^+ e^-)\tri$ processes, 
in order to evidentiate possible effects due to the exchange of a hypothetical low-mass boson,
the so-called X17.  These processes are studied for energies of the incident
beams in the range 18-30 MeV, in order to have a sufficient energy to produce such a boson,
whose mass is estimated to be around 17 MeV. 
We have first analyzed the reactions as a purely electromagnetic processes,
in the context of a state-of-the-art approach
to nuclear strong-interaction dynamics and nuclear electromagnetic currents,
the latter derived from $\chi$EFT. The initial $2+1$ scattering-state and trinucleon
bound-state wave functions have been obtained using the HH method. Therefore, the comparison with
accurate data for the cross sections of these processes could be already very useful (also in case of no observation of the X17), 
being a stringent test of our current knowledge of the strong and EM interactions
in nuclear systems. 

Next, we have investigated how the exchange of a hypothetical low-mass boson would impact the cross sections
for such processes. We have considered several possibilities, that this boson be either a scalar, pseudoscalar, vector, or
axial particle. The $A=3$ system does not present any excited-state structure, however the study of the cross section
and the eventual observation of a peak in the  $e^+$-$e^-$  angular distribution could be very instructive.
First of all, the experiment appears to be well feasible, since a deuteron target is easily manageable.
Varying the energy of the incident beam, under a certain threshold no peak should be observed, then the
variation of the position of the peak with increasing energy would give important constraints on the mass
of the X17. The variation of the height of the peak with energy would be also very instructive
on the nature of X17. Finally, the comparison of the $d(p,e^+ e^-)\het$ and $d(n,e^+ e^-)\tri$ cross
sections could give information on the coupling of X17 with protons and neutrons separately. Although
there is not much sensitivity to verify the alleged proto-phobicity in the case of a vector or pseudoscalar X17,
in the axial case the peak for the $n + d$ process is almost twice higher than in the $p + d$ case.

In the present treatment, the X17-nucleon interaction has been considered only at LO (with inclusion of some NLO
contributions) for the sake of simplicity. However, it can be extended 
to include higher order contributions in the framework of $\chi$EFT, as well. Work in this direction is in progress.
Moreover, we have limited ourselves to study the case of emission at identical polar angles $\theta=\theta'$.
Clearly further information about the nature of X17 and its couplings
with protons and neutrons could be achieved by detecting the leptons with
a large angular acceptance, useful also for increasing the statistics of the sub-dominant X17 channel. 

Finally, we comment about the feasibility of an experimental study of these two reactions.
  A nucleon beam with energy in the range 18-30 MeV impinging on a deuteron target can produce, in
  addition to elastic scattering, the following processes:
  1) spallation, via the breakup process $d(p/n,pn)p/n$ ($\sigma_{\mathrm {spall}}\sim 10^2$ mb~\cite{Kievsky:2000eb}),
  2) radiative capture, via the process $d(p/n, \gamma)\het/\tri$ ($\sigma_{\mathrm {capt}}~\sim 10$ $\mu$b, see Table~\ref{tab:totxs}),
  3) ``standard'' electromagnetic IPC, i.e. the process  $d(p/n, e^+e^-)\het/\tri$ ($\sigma_{\mathrm {IPC}}~\sim 10^{-2}$ $\mu$b, see Table~\ref{tab:totxs}),
  and 4) X17 electron-positron pairs (we have estimated that the corresponding cross section can be of the 
  order of $\sigma_{\mathrm {X17}}~\sim 10^{-3}\div 10^{-4}$ $\mu$b). Therefore, an experimental apparatus will have to deal 
  with the relatively high rate of processes 1) and 2) with respect to the X17 production,
  in particular with the neutrons produced by the spallation 
  ($\sigma_{\mathrm {spall}}/\sigma_{\mathrm {X17}}\approx 10^{8}\div10^9$)
  and the $\gamma$'s produced by the capture   ($\sigma_{\mathrm {capt}}/\sigma_{\mathrm {X17}}\approx 10^{4}\div10^5$).
  This suggests the use of a ''light'' detector
  to minimise its sensitivity to photons produced by either the latter capture reaction or induced by $(n, n'\gamma)$ processes.
  The detection of protons can be in principle managed exploiting their short range and/or their low velocity with respect to $e^+e^-$. 
  A light detector is also useful to minimize the external pair creation (EPC) events due to the interaction of
  photons with the material surrounding the target.
  In any case, EPC pairs are mainly produced at small relative angles, far away from the X17
  signal region. Instead, the ``standard'' IPC pairs have a distribution with a much longer tail towards
  large angles. The effect of this irreducible background ($\sigma_{\mathrm {X17}}/\sigma_{\mathrm {IPC}}\approx 10^{-1}\div 10^{-2}$)
  can be only limited by minimizing the broadening of the X17 peak due to electron and positron multiple scattering
  on the materials surrounding the ${}^2$H target. In this regard, the possibility of measuring the $d(n/p,e^+e^-)\het/\tri$
  processes has been considered eventually exploiting the detector designed to study the $\het(n,e^+e^-)\heq$
  reaction~\cite{universe10070285,Gustavino:2024wgb}.



%

\acknowledgments
We gratefully acknowledge Rocco Schiavilla for useful discussions and the support of the INFN-Pisa computing center.

\appendix
\section{The scalar current}
\label{app:a}
  In this appendix, we give some more detail on the derivation of the nuclear scalar current induced by
  the X17-quark interaction. The QCD+external fields Lagrangian can be written in general as
  \begin{eqnarray}\label{eq:Lq}
  \mathcal{L}_{q}(x)&=&\mathcal{L}_{q}^{0}(x)+\overline{q}(x)\,\gamma^\mu \left[v_\mu(x)+\frac{1}{3}\,v_\mu^{s}(x)\right] q(x)\nonumber\\
  &+&\overline{q}(x)\gamma^\mu\gamma^5 \,a_\mu(x) ]q(x)\ , \nonumber \\
  &-&\overline{q}(x)[ s(x)-i\gamma^5 \,p(x) ]q(x)\ , 
 \end{eqnarray}
  where $\mathcal{L}_q^{0}(x)$ is the Lagrangian for massless quarks,
  $q(x)$ is the two-component spinor
  \begin{equation}
  q(x)=\left[\!\begin{array}{c} 
         u(x)\\
         d(x)
         \end{array}\!\right]\ ,
  \end{equation}
  and $u(x)$ and $d(x)$ are the up and down  quark fields,  respectively.
  The quantities  $s(x)$, $p(x)$, $v_\mu^{(s)}(x)$, $v_\mu(x)$, and $a_\mu(x)$
  are $2\times2$ matrices in the flavour space describing the interaction with external fields
  (and also the quark mass term, seen in this formalism as an external perturbation
  breaking the chiral symmetry).

  In this appendix we focus on the scalar term $s(x)$,  
  so we can consider $p=v_\mu=v_\mu^{(s)}=a_\mu=0$ (for a full discussion, see Ref.~\cite{Viviani:2021stx}).
  This quantity can be decomposed as $s(x)=\sum_{i=0,3} \tau_i s_i(x)$, where $\tau_0$ is the identity matrix and
  $\tau_i$, $i=1,\ldots,3$  Pauli matrices. 
  Considering the X17-quark coupling given in Eq.~(\ref{eq:LqX1}) for the scalar case ($c=S$),
  we can identify 
  \begin{eqnarray}
    s_0(x)&=&-e \, {m_q\over\Lambda_S}  \varepsilon_0^S\, X(x)\ , \label{eq:s0}\\
    s_z(x)&=& -e\,  {m_q\over\Lambda_S}  \varepsilon_z^S\, X(x)\ ,\label{eq:s3}
  \end{eqnarray}
  where $\varepsilon_0^S$ and $\varepsilon_z^S$ are the coupling constants defined in Eqs.~(\ref{eq:eps_p_s})
  and~(\ref{eq:eps_m_s}), respectively. Clearly in our case $s_x(x)=s_y(x)=0$.
  
  The chiral Lagrangian is constructed in terms of the doublet of nucleon fields $N(x)$,
  the triplet of pion fields ${\bm\pi}(x)$ (a vector in isospin space), and the external field $s(x)$, using general principles
  and symmetries (in particular the chiral symmetry), see for example Refs.~\cite{Gasser:1983yg,Fettes:2000gb}.
  At lowest order, $s(x)$ enters the second-order pion-nucleon chiral Lagrangian
  through the quantity $\chi_+$,
  \begin{equation}
  {\cal L}_{\pi N}^{(2)}  = \overline{N}(x)\Bigl(c_1\langle \chi_+\rangle + c_5\hat\chi_+
    +\cdots \Bigr)N(x)\ , \label{eq:piN2}
  \end{equation}
  where  
 \begin{eqnarray}
    N(x) &=&\left[\begin{array}{c} 
         N_p(x)\\
         N_n(x)
         \end{array}\right]\ ,\label{eq:N} \\
    \chi_+ &=& u^\dag \chi \,u^\dag + u\, \chi^\dag u\ ,\qquad u=\sqrt{U}\ ,\label{eq:chipm}\\
     U&=&1+{i\over f_\pi} {\bm \tau} \cdot {\bm \pi}(x)-{1\over 2f_\pi^2}\, {\bm \pi}(x)^{2}+ \cdots
  \ , \label{eq:U}
 \end{eqnarray}
  with $\chi = 2\,B_c\,s(x)$, while $N_p(x)$ and $N_n(x)$ are the proton and neutron fields, respectively.
  Above $\langle\cdots\rangle$ denotes a trace over isospin,
  $\hat\chi_+=\chi_+-\langle\chi_+\rangle/2$,
  and the various constants $f_\pi$, $c_1$, $c_5$, $B_c$ are the so-called low-energy constants (LECs),
  usually determined by comparing with some experimental data (for example, 
  $f_\pi$ is the pion decay constants, $c_1$ and $B_c$ are related to the pion mass, etc.). Expanding
  ${\cal L}_{\pi N}^{(2)} $  in terms of the pion field and just retaining the zero-order term,
  we obtain the lowest nucleon-X17 interaction term:
  \begin{eqnarray}
    {\cal L}^S_{N,X}(x)&=& \overline{N}(x) \Bigl[ -8\,B_c\, c_1\, s_0(x) \nonumber \\
      && \qquad -4\, B_c \,c_5\, s_z(x) \, \tau_z \Bigr] N(x) X(x) \ .\label{eq:XN}
  \end{eqnarray}  
  Substituting the expressions of $s_0$ and $s_z$, one arrives at Eq.(\ref{eq:e4s}),
  with  the coupling constants given in Eq.~(\ref{eq:cpm}).

  The Hamiltonian $H_I$ is derived from this Lagrangian term 
  as detailed, for example, in Ref.~\cite{Baroni:2015uza}
  [in practice, $H_I=-\int d\bmx\, {\cal L}^S_{N,X}(t=0,\bmx)$].
  The amplitude is simply given by a diagram where a nucleon in momentum-spin-isospin state
  $|\bmp,s,t\rangle$ emits an X17 with momentum $\bmq$, changing its state to $|\bmp',s',t'\rangle$. 
  This is easily calculated as
  \begin{eqnarray}
    \langle\bmp',s',t'; \bmq|H_I|\bmp,s,t\rangle&=& e (\eta_0^S+\eta_z^S\tau_z)_{t',t}
    \overline u(\bmp',s') u(\bmp,s)\nonumber \\
    &&\times \delta_{s',s}\delta_{\bmp,\bmp'+\bmq} \ ,\label{eq:MEq}
  \end{eqnarray}
  where $u(\bmp,s)$ are Dirac four spinors. We then adopt the non-relativistic approximation
  $\overline u(\bmp',s) u(\bmp,s)\approx 1$. The amplitude for a collection of $A$ nucleons
  in $r$-space, $\langle\bmr_1',\ldots,\bmr_A'|H_I|\bmr_1,\ldots,\bmr_A\rangle$,
  can be obtained by inserting complete set of momentum states and using Eq.~(\ref{eq:MEq}).
  In this way, the operators given in Eqs.~(\ref{eq:JXS}) and~(\ref{eq:e94}) are obtained.

  An analogous procedure (see, for more details, Ref.~\cite{Viviani:2021stx}) allows for the
  derivation of the other operators given in Eqs.~(\ref{eq:JXP})--(\ref{eq:JXA2}).


\section{The five-fold cross section in region B}
\label{app:b}

In this appendix, we discuss in more detail the reasons of
approximating the five-fold cross section in region B as in 
Eq.~(\ref{eq:regB}). We start from the general expression for the
cross section given in Eq.~(\ref{eq:e13}). Let us consider the quantity
\begin{eqnarray}
  C_X &=& {\varepsilon_e^c R_X(\epsilon,\hat{\bmk}, \hat{\bmk}^\prime)D_X^*/Q^2+c.c.\over |D_X|^2}\nonumber\\
  &+& {(\varepsilon_e^c)^2  R_{XX}(\epsilon,\hat{\bmk}, \hat{\bmk}^\prime)\over |D_X|^2}\label{eq:CX1}
\end{eqnarray}
representing the contribution of the X17 to the cross section. 
Due to the very small value of $\Gamma_X$, see Eq.~(\ref{eq:gammaX}),
$|D_X|^2$ can be approximated as in Eq.~(\ref{eq:eeee}), where $\epsilon_i$
are the (two) values of $\epsilon$ (the electron energy)
where $Q^2-M_X^2=0$. Substituting  Eq.~(\ref{eq:eeee}) in the expression of $C_X$ and
written $\Gamma_X= \alpha(\varepsilon_e^c)^2 \Gamma_X^{(0)}$, where $\Gamma_X^{(0)}\approx M_X$, we
obtain
\begin{eqnarray}
  C_X &=& \sum_{i=1,2} \Bigl[ {2\varepsilon_e^c \Im\bigl [R_X(\epsilon,\hat{\bmk}, \hat{\bmk}^\prime)\bigr]
      \alpha M_X \Gamma_X^{(0)} \over Q^2} \nonumber \\
    && \qquad +R_{XX}(\epsilon,\hat{\bmk}, \hat{\bmk}^\prime)\Bigr] \gamma_i \delta(\epsilon-\epsilon_i) \ .
    \label{eq:CX2}
\end{eqnarray}
We remember that  $R_{X}$ ($R_{XX}$) depends linearly (quadratically) on the X17-hadron coupling constants $\eta^c_{0,z}$.
However, the factor multiplying $\Im[R_{X}]$ is small, in fact
\begin{equation}
  {2\varepsilon_e^c \alpha M_X \Gamma_X^{(0)} \over Q^2}
  = 2\varepsilon_e^c \alpha { \Gamma_X^{(0)} \over M_X}\sim 10^{-5}\ ,
\end{equation}
since $\varepsilon_e^c\sim 10^{-3}$ and the $\delta(\epsilon-\epsilon_i)$ imposes $Q^2=M_X^2$.
Consequently, we have found that the contribution of the term $\Im[R_X]$
can be always neglected in comparison to $R_{XX}$ and Eq.~(\ref{eq:regB}) follows.
This is true if the $|\eta^c_{0,z}|$ are not smaller than $10^{-3}$, as for the cases discussed
in this paper. In any case, in the numerical calculations both terms are taken into account.

\bibliographystyle{apsrev4-1}
\bibliography{x17-pd}{}
\end{document}